\def\wig#1{\mathrel{\hbox{\hbox to 0pt{%
          \lower.5ex\hbox{$\sim$}\hss}\raise.4ex\hbox{$#1$}}}}

\def\v1n{{\cal U}^N_1}

\def\sss{\scriptscriptstyle}
\def\phih2h2{\phi_{\sss {\rm H_2-H_2}}}
\def\phh2{\phi_{\sss {\rm H-H_2}}}

\def\Teff{T_{\rm eff}}
\def\sqr#1#2{{\vcenter{\vbox{\hrule height.#2pt 
  \hbox{\vrule width.#2pt height#1pt \kern#1pt 
  \vrule width.#2pt} 
  \hrule height.#2pt}}}}

\def\wig#1{\mathrel{\hbox{\hbox to 0pt{%
          \lower.5ex\hbox{$\sim$}\hss}\raise.4ex\hbox{$#1$}}}}

\documentclass[preprint2]{aastex}

\shortauthors{Saumon et al.}
\shorttitle{Molecular abundances in Gl 229B}
\slugcomment{DRAFT:  DO NOT CIRCULATE!}

\begin{document}

\title{Molecular Abundances in the Atmosphere of the T Dwarf Gl 229B}

\author{D. Saumon} 
\affil{Department of Physics and Astronomy, Vanderbilt University, Nashville, TN 37235} 
\email{dsaumon@cactus.phy.vanderbilt.edu}
\author{T.R. Geballe}
\affil{Gemini North Observatory, 670 North A'ohoku Place, Hilo, HI 96720}
\author{S.K. Leggett}
\affil{Joint Astronomy Center, 660 North A'ohoku Place, Hilo, HI 96720}
\author{M.S. Marley}
\affil{Department of Astronomy, New Mexico State University, Las Cruces, NM 88003}
\author{R.S. Freedman}
\affil{The Space Physics Research Institute, NASA Ames Research Center} 
\affil{Moffett Field, CA 94035}
\author{K. Lodders, B. Fegley, Jr.}
\affil{Planetary Chemistry Laboratory, Department of Earth \& Planetary Science}
\affil{Washington University, St-Louis, MO 63130}
\and
\author{S.K. Sengupta}
\affil{Department of Physics and Astronomy, Vanderbilt University, Nashville, TN 37235} 

\begin{abstract}
We present new, high resolution, infrared spectra of the T dwarf
Gliese 229B in the $J$, $H$, and $K$ bandpasses.  We analyze each of these as well as
previously published spectra to determine its 
metallicity and the abundances of NH$_3$ and CO in terms of the surface
gravity of Gl 229B,  which remains poorly constrained.   The metallicity increases with increasing
gravity and is below the solar value unless Gl 229B is a high-gravity brown dwarf with
$\log g \,({\rm cgs}) \approx 5.5$.  The NH$_3$ abundance is determined
from both the $H$ and the $K$ band spectra which probe two different levels in the
atmosphere.  We find that the abundance from the $K$ band data is well below that expected
from chemical equilibrium, which we interpret as strong evidence for dynamical transport 
of NH$_3$ in the atmosphere.  This is consistent with the previous detection of CO and 
provides additional constraints on the dynamics of the atmosphere of this T dwarf.
\end{abstract}


\section{Introduction}

Gliese 229B is not only the first brown dwarf recognized as genuine \citep{naka95,oppen95},
but it is also the brightest and best-studied T dwarf known.  
With an effective temperature of $\Teff \sim 950\,$K, it
lies squarely between the latest L dwarfs ($\Teff \sim 1500\,$K, \citet{kirk98}) and 
the giant planets 
of the solar system ($\Teff \sim 100\,$K).  Indeed, its near infrared spectrum  shows the 
strong
H$_2$O absorption bands characteristic of very-low mass stars and the strong CH$_4$ bands
seen in the spectra of Jupiter, Saturn and Titan.  The transitional nature of the 
spectrum of Gl 229B is remarkable and hints at the spectral appearance of extrasolar
giant planets which have effective temperatures in the range 200 -- 1600$\,$K \citep{guillot99}.

A wealth of data on Gl 229B has accumulated since its discovery five years ago.
Broad band photometry from $R$ through $N$  and an accurate parallax \citep{matth96,
golim98, legg99, hipparcos}
allow an accurate determination of its bolometric luminosity. 
Spectroscopic observations \citep{oppen98,geb96,schultz}
covering the range from  0.8 to 5.0$\,\mu$m 
have revealed a very rapidly declining flux shortward of 1$\,\mu$m, the unmistakable presence 
of CH$_4$, H$_2$O, and Cs, and demonstrated the 
{\it absence} of the  CrH, FeH, VO and TiO features characteristic of late M and early 
L dwarfs.  Finally, Noll, Geballe \& Marley (1997) and \citet{oppen98} have
detected CO with an abundance well above the value predicted by chemical equilibrium, a 
phenomenon also seen in the atmosphere of Jupiter.
 
Model spectra for Gl 229B \citep{mar96,allard96,tsuji96b}
reproduce the overall energy distribution fairly well and all agree  that 
1) $\Teff \sim 950\,$K, 2) compared to gaseous molecular opacity, the dust opacity 
is small if not negligible in the infrared,
3) the gravity of Gl 229B is poorly constrained at present.
The rapid decline of the flux at wavelengths shortward of 1$\,\mu$m is interpreted
as caused by an absorbing haze of complex hydrocarbons (Griffith, Yelle \& Marley 1998) or 
alternatively by the pressure-broadened red wing of the K I resonance doublet  at 0.77$\,\mu$m
(Tsuji, Ohnaka \& Aoki 1999; Burrows, Marley \& Sharp 1999).

In this paper, we present new high-resolution spectra in the
$J$, $H$, and $K$ bands. With the inclusion of the ``red'' spectrum of \citet{oppen98},
we analyze each part of the spectrum separately to obtain independent measures of
the H$_2$O abundance of Gl 229B -- broadly interpreted as the metallicity index --  to
detect for the first time the presence of NH$_3$ in its spectrum, and to estimate the
NH$_3$ abundance at two different depths in the atmosphere.  
Our results are expressed in terms of the
surface gravity which cannot be determined from the data presented here.
Nevertheless, we identify a reduced set of acceptable combinations of $\Teff$ and gravity,
using the observed bolometric luminosity of Gl 229B \citep{legg99}.

The observations and the near infrared spectra are presented in \S2. Section 3 shows
how an accurate parallax, a well-sampled spectral energy distribution and evolutionary 
models 
greatly reduce the possible range of combinations of $\Teff$ and gravity
without having to resort to spectrum fitting.  The synthetic spectrum calculation and 
our method of analysis are described in \S4.  The results concerning several molecules
of interest which are at least potentially detectable are presented in \S5, followed
by a discussion in \S6.  Finally, a summary of the results and directions for future 
study are given in \S7.

\section{Spectroscopic observations}

Spectra of Gl~229B in selected narrow intervals in the $J$, $H$, and $K$ windows
were obtained at the 3.8~m United Kingdom Infrared Telescope (UKIRT) in
1998 January, using the facility spectrometer CGS4 \citep{mountain}
and its 150 l/mm grating. Details of the observations are provided in
Table 1. These are among the highest resolution spectra obtained of any
T dwarf.

The spectra were obtained in the standard stare/nod mode with the
$1.2^{\prime\prime}$ wide slit of the spectrometer oriented at a position angle of
$45^\circ$, nearly perpendicular to the line connecting Gl 229A and Gl 229B.
The southward-going diffraction spike of Gl 229A together with scattered
light from that star, which is 10 magnitudes brighter than Gl 229B,
contaminated the array rows near and to the southwest of those containing
the spectrum of Gl 229B. The contamination on the Gl 229B rows was
determined by interpolation and was subtracted; typically it was
comparable or somewhat smaller than the signal from Gl~229B. In order to
remove telluric absorption features, spectra of the A0V star BS~1849 were
measured just prior to Gl~229B. In all cases the match in airmasses was
better than five percent and hence in the ratioed spectra residual
telluric features are small compared to the noise level. Wavelength
calibration was achieved by observations of arc lamps and is in all cases
better than one part in $10^{4}$ ($2\sigma$).

The spectra shown in this paper have been slightly smoothed, so that the
resolving powers are lower than those in Table 1 by approximately 25
percent. They also have been rebinned to facilitate coadding like spectra
and joining adjacent spectral regions. The error bars can be judged by
the point-to-point variations in featureless portions of the spectra, the
signal-to-noise ratios at the continuum peaks are
approximately 40 in the $K$ band, 25 in the $H$ band, and 30 in the $J$ band.
The flux calibration of each spectrum is approximate as no attempt was made
to match the photometry of Gl 229B.

While we identify the spectra by their corresponding standard photometric
infrared bandpass, their wavelength coverages are much narrower than the $JHK$ filters
and typically corresponds to the peak flux of Gl 229B in each bandpass.

In the $J$ band spectrum (Fig. 1), nearly all features are caused by H$_2$O.  The
short wavelength end of the spectrum shows the red side of a CH$_4$ band, which
is responsible for the features seen shortward of $\sim 1.215\,\mu$m.  Two lines of
neutral potassium are easily detected near $1.25\,\mu$m.  No other lines of alkali
metals fall within the wavelength coverage of our observations.
The $H$ band spectrum (Fig. 2) is relatively rich in molecular opacity sources.  All
the features seen in the spectrum are either due to H$_2$O ($\lambda \wig< 1.59\,\mu$m)
or part of a very strong CH$_4$ absorption band ($\lambda \wig> 1.6\,\mu$m).  Features seen
between 1.59 and 1.6$\,\mu$m cannot presently  be ascribed with certainty and are due 
to either H$_2$O or CH$_4$.  While the opacities of NH$_3$ and H$_2$S are not 
negligible in this
part of the spectrum, neither molecule forms distinctive spectral features.  Their presence
cannot be directly ascertained from these data, mainly because their opacity is weaker than
that of H$_2$O and CH$_4$ and because of significant pressure broadening (see \S 5.3).
The $K$ band flux emerges in an opacity window between a strong H$_2$O band and a strong
CH$_4$ band (Fig. 3).  Spectral features are caused by H$_2$O at shorter wavelengths 
($\lambda \le 2.11
\,\mu$m) and by CH$_4$ at longer wavelengths ($\lambda \ge 2.105\,\mu$m).  Several 
distinctive
features of NH$_3$ are expected at the blue end of this spectrum and models predict a single
absorption feature of H$_2$S at $2.1084\,\mu$m.

All features seen in Figures 1 -- 3 are unresolved blends of numerous molecular transitions.  
A spectral resolution at least 10 times higher would be required to resolve the intrinsic
structure of the spectrum of Gl 229B.

\section{Effective temperature and gravity}

While synthetic spectra have been fairly successful at reproducing the unusual spectrum
of Gl 229B \citep{mar96,allard96,tsuji96b}, the entire spectral energy distribution
has not yet been modeled satisfactorily.  Limitations in the opacity databases  are 
partly responsible for the remaining discrepancies between synthetic spectra and the data
(see \S4.2).  These shortcomings have impeded the determination of $\Teff$ and
of the gravity $g$ in particular.  On the other hand, the bolometric luminosity of Gl 229B is
now well determined.  Combining spectroscopic and photometric data from 0.82 to
10$\,\mu$m with the parallax, \citet{matth96} found $L=6.4 \times 10^{-6}\,L_\odot$. 
With new $JHKL^\prime$ photometry,
\citet{legg99} found $L=6.6 \pm 0.6 \times 10^{-6}\,L_\odot$.  A recalibration 
using the HST photometry of \citet{golim98} gives $L=6.2 \pm 0.55 \times 10^{-6}\,L_\odot$. 
Evolutionary models \citep{egp2}
allow us to find a family of $(\Teff,g)$ models with a given $L_{\rm bol}$.
Figure 4 shows the cooling tracks of solar metallicity brown dwarfs in terms of the surface 
parameters $\Teff$ 
and $g$.  Models with the bolometric luminosity of Gl 229B  fall within the
band running through the center of the figure.  Using a very conservative lower 
limit for the age of Gl 229A of 0.2$\,$Gyr \citep{naka95}, we find $\Teff = 950 \pm 80\,$K.
On the other hand, the gravity remains poorly determined with $\log g\,({\rm cm/s}^2) = 5
\pm 0.5$, corresponding
to a mass range of 0.015 -- 0.07$\,M_\odot$.  While the upper range is very
close to the lower main sequence mass limit, Gl 229B's status as a brown dwarf is 
secure.  A star at the edge of the main sequence would be much hotter with 
$\Teff \sim 1800\,$K; well outside of Fig. 4.  

In the remainder of this paper, the discussion focuses on three 
atmosphere models which span the range of allowed solutions (Table 2 and Fig. 4): 
$(\Teff\,({\rm K}),\log g\,({\rm cgs}))=$ (870, 4.5), (940, 5.0) and (1030, 5.5), which
we label models A, B, and C, respectively.  These constraints on $\Teff$ and $g$ from
cooling sequences are quite firm.  We find the same result, within the error bar on
$L_{\rm bol}$, from several cooling sequences which predate \citet{egp2}. The latter were
computed with different input physics such as the
equation of state and surface boundary conditions derived from grey and non-grey 
atmosphere models using several opacity tabulations.

In section 5, we  show that these three models can fit the spectra only if they have different 
metallicities, ranging from [M/H]$=-0.1$ to $-0.5$.  The evolution of brown dwarfs is
sensitive to the metallicity through the atmospheric opacity which controls the rate of
cooling.  We find that for
the range of interest here, the effect of a reduced metallicity on our determination of
$\Teff$, $g$, and the cooling age is smaller but comparable to
that of the uncertainty on the value of $L_{\rm bol}$. We choose to ignore it for simplicity.

\section{Method of analysis}

\subsection{Model atmospheres and spectra}

Our analysis is based on the atmosphere models of brown dwarfs and extrasolar giant 
planets described in \citet{egp2}.
Briefly, the atmospheres are in radiative/convective equilibrium and the equation of 
radiative transfer is solved with the k-coefficient method.  The
chemical equilibrium is treated as in \citet{egp2}.
Gas phase opacities include Rayleigh scattering, the collision-induced 
opacity of H$_2$ and the molecular opacities of H$_2$O, CH$_4$, NH$_3$, H$_2$S, 
PH$_3$, and CO,
as well as the continuum opacities of H$^-$ and H$_2^-$.  The 
molecular line opacity database is described in more detail in \S4.2.
Atomic line opacity is not included.  
Because of the relatively large gravity of Gl 229B,
pressure broadening of the molecular lines  plays an important role in determining
the $(T,P)$ profile of the atmosphere and in shaping the spectrum.  The line-by-line
broadening theory we use is described in \citet{egp2}.
The strong continuum opacity source responsible for the rapid decrease of the flux of
Gl 229B shortward of 1.1$\,\mu$m is included following the haze model of \citet{gym98}.
Details of the haze model and of our fitting procedure are given in \S5.1.
The $(T,P)$ structures of these atmosphere  models
are shown in Fig. 5.  The profiles intersect each other 
because both $\Teff$ and the gravity vary between 
the models.  The inflexion point at $\log T \sim 3.25$ signals the top of the convection zone.

Using the same monochromatic opacities used to compute the k-coefficients,
high-resolution synthetic spectra are generated from the atmospheric structures 
by solving the radiative transfer equation with the Feautrier method on
a frequency grid with $\Delta \nu=0.1\,$cm$^{-1}$.  Spectra with resolution lower than
$\Delta \nu \wig> 1\,$cm$^{-1}$ can then be generated for comparison with data.  

An unusual aspect of T dwarf atmospheres is the great variation of the 
opacity with 
wavelength.  These atmospheres are strongly non-grey and the near infrared spectrum 
is 
sculpted by strong absorption bands of H$_2$O and CH$_4$.  Most of the flux emerges in
a small number of relatively transparent opacity windows.  The concept of photosphere 
becomes rather useless since the level at which the spectrum is formed depends strongly
on the wavelength.  Figure 6 shows the depth of the ``photosphere'' $(\tau_\nu=2/3)$ in
both temperature and pressure as a function of wavelength for model B ($\Teff=940\,$K,
$\log g=5$).  
In the $Z$, $J$, and $H$ bands, and, to a lesser extent, in the $K$ and $M$ bands, the 
atmosphere is very transparent and can be probed to great depths.
For $\lambda \wig> 25\,\mu$m,  the spectral
energy distribution approaches a Planck function with $T \sim 500\,$K. 

Figure 6 reveals that spectroscopy between 0.8 and 12$\,\mu$m can probe the atmosphere from
$T \sim 500\,$K down to a depth where $T \sim 1600\,$K,  corresponding to a range of 6 
pressure scale heights! This provides an exceptional opportunity to study the physics of the 
atmosphere of a brown dwarf over an extended vertical range.
The top of the convection zone for model B (located at $T=1860\,$K) is
below the ``photosphere'' at all wavelengths and is not directly observable, however.

\subsection{Limitations to this study}

Given the range of acceptable values of $\Teff$ and $g$ (\S3), we can determine 
the metallicity of Gl 229B and the abundance
of several key molecules by fitting synthetic spectra to the observations, for each of the
three models.  The precision of
our results is determined by the reliability of the models, the noise level in the data and,
most significantly, by the limitations of the molecular line lists used to compute their
opacities.  The latter point requires a detailed discussion.

The opacities of CH$_4$ and NH$_3$ are computed from line lists obtained by combining the HITRAN
\citep{hitran} and GEISA \citep{geisa} databases, which are complemented with recent laboratory 
measurements  and theoretical calculations.  Further details are provided in \citet{egp2}.
The resulting line lists for these two molecules are very nearly complete for $T<300\,$K,
and their degree of completeness decreases rapidly at higher temperatures where absorption
from excited level become important.  Furthermore, the line list
of CH$_4$ is limited to $\lambda > 1.584\,\mu$m.  We extend the CH$_4$ opacity to shorter wavelengths
with the laboratory measurements of \citet{strong} which provide the absorption coefficient 
averaged over intervals of 5$\,$cm$^{-1}$ between 1 and 5$\,\mu$m at $T=300\,$K. 
We use \citet{strong} opacities for
$1 < \lambda < 1.82\,\mu$m and the line list for $\lambda > 1.82\,\mu$m.  This puts the transition
from one tabulation to the other in a strong H$_2$O absorption band and obliterates any
discontinuity in CH$_4$ opacity at the transition.  For $\lambda < 1\,\mu$m, the tabulation 
of \citet{kark} 
gives the absorption coefficient of CH$_4$ determined from spectroscopic observations
of the giant planets at 0.0004$\,\mu$m intervals.  Because of the low temperatures found
in the atmospheres of giant planets, the Karkoschka CH$_4$ opacities are appropriate for
$T \wig< 200\,$K.  To our knowledge, this compilation of NH$_3$ and CH$_4$ opacity 
is the most complete presently available.

As can be seen in Figure 6, the temperature in
the atmosphere of Gl 229B is everywhere greater than 300$\,$K.  For CH$_4$, we compute
temperature-dependent line opacity (which is incomplete above 300$\,$K) from the
population of excited levels determined by the Boltzmann formula for $\lambda > 1.82\,\mu$m,
and use temperature independent opacity at shorter wavelengths.  While the synthetic spectra
computed reproduce the fundamental band of CH$_4$ very well (centered at $\lambda= 3.4\,\mu$m),
the match with the 1.6 and 2.3$\,\mu$m bands is rather poor.  Even though CH$_4$ is a very
prominent molecule in the spectrum of Gl 229B, the current knowledge of its opacity is not
adequate for a quantitative analysis of its spectral signature.  For this reason, we have 
essentially 
ignored the regions in our spectra where CH$_4$ is prominent.  Unfortunately, this prevents
us from estimating the abundance of CH$_4$ -- and therefore of carbon -- in Gl 229B.

Ammonia shows significant absorption in both the $H$ and $K$ band spectra.  The line list
for NH$_3$ starts at $\lambda > 1.415\,\mu$m. 
Because the line list  does not include transitions from highly excited levels which occur
at $T > 300\,$K, the NH$_3$ opacity we compute at a given wavelength is strictly a lower limit
to the actual opacity.

Except for the collision-induced absorption  by H$_2$, the most important molecular absorber 
in Gl 229B is H$_2$O, for which the opacity is now relatively well understood.  We use the most
recent and most extensive ab initio
line list (\citet{partridge}, $3 \times 10^8$ transitions).  This line list 
is essentially complete for $T \wig< 3000\,$K.
As a demonstration of the equality of this database, we find that
the H$_2$O features computed with this line list
correspond extremely well in frequency with the observed features of Gl 229B ({\it e.g.} Figs. 7 to 9).  
However, we find noticeable discrepancies in the relative strengths of H$_2$O features which
we attribute to the calculated oscillator strength of the transitions (\S 5.1).  This effect can 
also be
seen in Fig. 1c of \citet{gym98}.    At high resolution, the distribution of molecular 
transitions in frequency and strength is nearly random, and the inaccuracies in oscillator
strengths we have found limit the accuracy of model fitting in a fashion similar to noise.
This ``opacity noise'' is at least as significant as the noise intrinsic to our data.

\section{Results from fitting the spectra}

We have constructed a grid of synthetic spectra for the three models shown in Fig. 4 with
metallicity $-0.7 \le {\rm [M/H]} \le 0.1$ in steps of 0.1.  We use these modeled spectra to fit
four distinct spectral regions (the ``red'', $J$, $H$, and $K$ spectra) separately to
determine the metallicity as a function of gravity with an internal precision of $\pm 0.1$
dex.  For the purpose of fitting the data, the synthetic 
spectra were renormalized to the observed flux at a selected wavelength in each spectral 
region.

The model spectra show some distortions in the overall shape of the spectrum which are
probably due to remaining uncertainties in the $(P,T)$ profile of the atmosphere, the
inadequate CH$_4$ opacities, and possible effects of dust opacity.
Considering the additional problems with the strength of the H$_2$O features, 
we elected to do  all fits ``by eye,'' except where otherwise noted.  
We discuss the fitting of each spectral interval
below.  In the interest of brevity, we present a detailed discussion of fits
obtained only with the model of intermediate gravity (model B).
The best fits obtained with models A and C are very nearly identical to
those with model B.  The results are summarized in Table 2. 

\subsection{The ``red'' spectrum}

The ``red'' spectrum extends from 0.83 to $\sim 1\,\mu$m and is formed deep in the
atmosphere where $1100 < T < 1500\,$K.
The spectra of \citet{schultz} and \citet{oppen98} reveal
two lines of Cs I (at 0.852 and 0.894$\,\mu$m) and a strong H$_2$O band but {\it not} 
the bands of TiO and VO common to late M dwarfs and early L dwarfs (Fig. 7).  
Refractory elements, such as Ti, Fe, V, Ca,
and Cr, are expected to be bound in condensed compounds in a low-temperature 
atmosphere such as that of Gl 229B and therefore are not available to form molecular
bands \citep{fl96,mar96,bs99}.  
With the exception of the strong, unidentified feature at 0.9874$\,\mu$m,  all features
between 0.89 and 1.0$\,\mu$m can be attributed to H$_2$O. An overlap of a weak band
of H$_2$O and a weaker CH$_4$ band causes the small depression
at 0.894$\,\mu$m which was tentatively tentatively attributed to CH$_4$ by \citet{oppen98} 
and \citet{schultz}.  Features below 0.89$\,\mu$m cannot be identified at present.
The two Cs I lines are not included in our model. 

The flux from Gl 229B is also observed to 
decrease very rapidly toward shorter wavelengths \citep{schultz,oppen98,golim98}, which, 
in the absence of the 
strong TiO and VO bands, is evidently caused by  the presence of a missing source 
of opacity in the atmosphere. Spectra computed with molecular opacities only (but
excluding TiO, VO, FeH, etc) predict visible 
fluxes which are grossly overestimated \citep{gym98} but the detailed sequence of absorption 
features of the spectrum are well reproduced, indicating that the short wavelength 
flux is suppressed by a {\it smooth} opacity source.

We fit the red spectrum of Gl 229B  between 0.82 and 1.15$\,\mu$m to obtain the metallicity.
We have recalibrated the published spectrum \citep{oppen98,geb96,legg99} using the 
HST photometry of \citet{golim98}.  We model the continuum
opacity with a layer of condensates following the approach of \citet{gym98}.
Condensates are expected in the atmosphere of Gl 229B on the basis of
chemical equilibrium calculations \citep{lodders99,bs99}
and can provide the required opacity.  Alternatively, Tsuji, Ohnaka \& 
Aoki (1999) and Burrows, Marley, \& Sharp (1999) attribute this rapid decline to the
pressure broadened red wing of the 0.77$\,\mu$m K I resonance doublet. The first optical
spectrum of a T dwarf (SDSS 1624+0029) shows that the latter explanation is correct \citep{liebert00}.
The nature of this opacity source
is not very important for the determination of the metallicity, however, as long as the proper 
continuum opacity background is present in the calculation.

The dust opacity is computed with the Mie theory of scattering by spherical particles and is 
determined by the vertical distribution of the particles,
their grain size distribution, and the complex index of refraction of the condensate.
The cloud model of \citet{gym98} is described by 
1) the vertical density profile of the condensate, taken as:
$$n_{\sss d}= AP,$$
where $n_{\sss d}$ is the number density of condensed particles, $P$ is the 
ambient gas pressure,  and the cloud layer 
is bound by 
$$P_{\rm top} \le P \le 100\,{\rm bar};$$
2) the size distribution of the particles
$$f(d)={d \over d_0} \exp\Bigg[{\ln(d/d_0) \over \ln \sigma}\Bigg]^2,$$
where $d$ is the diameter of the particles; and
3) the complex index of refraction of the condensate
$$n(\lambda)=1.65 + in_i(\lambda).$$
The parameters $P_{\rm top}$, $A$, $d_0$, $\sigma$, the function $n_i(\lambda)$, and the
metallicity of the atmosphere are free parameters.

Such a multi-parameter fit of the observed spectrum is not unique.  Furthermore,
arbitrarily good fits of the ``continuum'' flux level
can be obtained by adjusting the imaginary part of the index of refraction since
its wavelength dependence is weakly constrained {\it a priori}.  Our results for the
three models are qualitatively similar to those of \citet{gym98}.  Typical values of
the fitted dust parameters   are $A=415\,$cm$^{-3}$bar$^{-1}$, $P_{\rm top}=0.5\,$bar, 
$d_0=0.21\,\mu$m, and $\sigma=1.3$, with the imaginary part of the
index of refraction decreasing from $n_i=1.70$ at 0.8$\,\mu$m to 0.01 at 1.10$\,\mu$m.
These parameters applied to model B with a metallicity of
[M/H]=$-0.3$ result in the fit shown in Figure 7.  

The metallicity 
of the atmosphere [M/H] is largely independent of the dust parameters, however, as
it is constrained by the amplitude of the features in the
H$_2$O band which we fit between 0.925 and 0.98$\,\mu$m.  A larger metallicity results in a 
larger amplitude of the features inside the band.  We obtain the same value of [M/H]
as long as a good fit of the ``continuum'' flux level is obtained, regardless of the particular
values of dust parameters.  The lower panel of Fig. 7 clearly shows differences  in
the relative strengths of the absorption features in the H$_2$O band between the
synthetic and observed spectra.  Similar differences also occur
in the $J$, $H$, and $K$ bands. The same differences are found for all three atmospheric
profiles and point to inaccuracies in the oscillator strength of the ab initio
line list of H$_2$O \citep{partridge}.  The metallicity is fitted to give the best overall
fit of these features with a precision of $\pm 0.1$ dex.  

Strictly speaking, this procedure gives the H$_2$O abundance, or [O/H] rather than [M/H].
For solar metallicity, the condensation of silicates deep in the atmosphere of Gl 229B will 
reduce the amount of
oxygen available to form H$_2$O by $\sim 15$\% \citep{fp88} which implies that
\begin{equation}
 {\rm [M/H]} = {\rm [O/H]} + 0.07.
\end{equation}
The correction, which decreases for subsolar metallicities, is smaller than our fitting uncertainty 
and will be ignored hereafter.

\subsection{$J$ band spectrum}

The $J$ band spectrum probes the most transparent window of the spectrum of Gl 229B and
is formed at great depth where $1500 \le T \le 1600\,$K  and $P \sim 30\,$bar (Fig. 6).
Our spectrum contains almost exclusively H$_2$O features, with the exception
of CH$_4$ absorption for $\lambda \wig< 1.215\,\mu$m and of two prominent K I lines
(Fig. 1).  Since we have elected to ignore CH$_4$ bands and our synthetic spectra do
not include alkali metal lines, we fit the $J$ spectrum between 1.215 and 1.298$\,\mu$m
to determine the metallicity from the depth of the H$_2$O absorption features.  
Figure 8 shows the effect of the metallicity on the spectrum (top panel) and our best 
fit (bottom panel) for model B. 
The flux level in the $J$ spectrum varies by a factor of 3 and our best fit shows distortions
in the general shape of the spectrum.  The distortions may be caused by a combination of
uncertainties in the
$(T,P)$ profile of the atmosphere, problems with the H$_2$O opacities or a small amount
of dust opacity in the infrared spectrum of Gl 229B (not modeled).
Since the fit is based on the depth of the features, we ignore these distortions and fit
the logarithm of the flux rather than the flux, as shown in Fig. 8.  
As in the red spectrum (\S5.1), we find a remarkable correspondence of spectral features
between the observed and modeled spectra although the model spectrum is somewhat less successful at
reproducing the relative strengths of the H$_2$O features.

\subsection{$H$ band spectrum}

While the $H$ band spectrum falls between the red side of a strong H$_2$O band
(for $\lambda \wig< 1.59\,\mu$m) and the blue side of a prominent CH$_4$ band (for
$\lambda \wig> 1.59\,\mu$m), NH$_3$ and H$_2$S have non-negligible opacity in this
wavelength interval which also includes a band of CO.   

The strong CH$_4$ band is responsible for the turnover of the flux at
1.59$\,\mu$m. In this wavelength range, the CH$_4$ opacity is described in our
calculation by the \citet{strong}
laboratory measurements which are restricted to $T=300\,$K.  As
a consequence, the CH$_4$ band comes in at 1.61$\,\mu$m in the synthetic spectra, which
results in strong departures of the mean flux level between data and models for 
$\lambda > 1.58\,\mu$m. This limits our analysis of the $H$ band spectrum to wavelengths
shorter than 1.58$\,\mu$m.  In this wavelength range, the spectrum is formed deep in the
atmosphere, where
$1200 < T < 1350\,$K and $P \sim 10\,$ bar (Fig. 6).  Within this spectral region, we 
determine the abundance of
NH$_3$ as a function of the metallicity, but cannot untangle the two.  We also comment
on the presence of H$_2$S and CO.

\subsubsection{Metallicity and ammonia}

While NH$_3$ absorption can significantly affect the slope of the spectrum
shortward of 1.56$\,\mu$m (Fig. 9), there is no
distinctive feature at this spectral resolution ($R=2100$) to provide an unambiguous 
detection
of this molecule.  The detection of CO at 4.7$\,\mu$m well above the equilibrium abundance
\citep{noll} suggests the possibility that NH$_3$ may also depart from its chemical equilibrium
abundance \citep{fl94}.  We therefore
vary the abundance of NH$_3$ by reducing its chemical equilibrium abundance -- {\it as
computed for a given metallicity} -- uniformly throughout the atmosphere by constant factor.
We found that reduction factors of 1, 0.5, 0.25, and 0 provide an adequate grid of
NH$_3$ abundances given the S/N ratio of the data and the residual problems with the H$_2$O
opacity.  

The effect of varying the metallicity on the $H$ band spectrum is shown in Fig. 10 for
model B.  All
spectra in Figure 10 were computed with the equilibrium abundance of NH$_3$.
In this case, we find a best fitting metallicity of [M/H]$=-0.3$.  Figures 9 and 10
show that varying the abundance of NH$_3$ and varying the metallicity have very similar
effects on the synthetic spectrum.  In the absence of any distinctive feature of NH$_3$,
it is not possible to determine both the metallicity and the NH$_3$ abundance separately.
For each value of [M/H], we can adjust the NH$_3$ abundance to obtain a good fit, higher
metallicities requiring lower NH$_3$ abundances.  The best fitting 
solutions are given in Table 2. These fits are nearly indistinguishable
from each other although the higher metallicity fits are marginally better.

It is in the $H$ band that we find the poorest match in the detailed features of the
data and the models.  While the fit is determined by matching the relative amplitudes of
the features, the two-parameter fit (metallicity and NH$_3$ abundance)
 we have performed amounts to little more than
fitting the slope of the spectrum.  As an internal check on our fitting procedure and
precision, we have have verified that our fits indeed have the same slope as the data
by plotting data and models at a very low spectral resolution which eliminates all absorption
features.

\subsubsection{Hydrogen sulfide}

Hydrogen sulfide (H$_2$S) has non-negligible opacity over most of the wavelength range of
our fit to the $H$ band.  Figure 11 shows two spectra computed with and without H$_2$S
opacity.  Its opacity is weaker than that of NH$_3$ however, and there are no distinctive
features which would allow a positive identification.  Since the H$_2$S features are
fairly uniformly distributed in strength and wavelength, fitting the spectrum with either
the chemical equilibrium abundance of H$_2$S or with no H$_2$S at all only has a 
small-to-negligible effect on the determination of the NH$_3$ abundance for a given metallicity 
(Table 2).  For the low gravity model A, the NH$_3$ abundance determined from spectra
without H$_2$S is about half of the value found with the chemical equilibrium abundance 
of H$_2$S. The difference decreases at higher gravities and is negligible for model C.

Chemical equilibrium calculations \citep{fl96,lodders99} indicate that H$_2$S is present in 
Gl 229B with an abundance essentially equal to the elemental abundance of sulfur in the
atmosphere (see \S 6.3 for further discussion).
Unfortunately, we are unable to ascertain the presence of H$_2$S in the atmosphere of Gl 229B
at present.

\subsubsection{Carbon monoxide}

The discovery of CO in the 4.7$\,\mu$m spectrum of Gl 229B with an abundance  about 3 orders of
magnitude higher than predicted by chemical equilibrium revealed the importance of dynamical 
processes in its atmosphere \citep{fl96,noll,griff99}.  The 4.7$\,\mu$m spectrum probes the 
atmosphere at the 2 -- 3 bar
level where $T \sim 900\,$K (Fig. 6).  At this level, chemical equilibrium calculations
predict a CO abundance of $X_{\rm CO} = 4.7 \times 10^{-8}$ 
while \citet{noll} found $5 \times 10^{-5} \wig< X_{\rm CO} \wig< 2 \times 10^{-4}$.
The second overtone band of CO falls within the $H$ band and, in principle,
could provide a determination of $X_{\rm CO}$ at a deeper level of the atmosphere, 
where $P \sim 14\,$ bar and $T \sim 1400\,$K (Fig. 6).
Figure 12 shows a comparison of the data with synthetic
spectra computed with various amounts of CO for model B with solar metallicity.  The first 
two spectra are computed with the 
chemical equilibrium abundance of CO ($X_{\rm CO}=4.90 \times 10^{-5}$ at 14$\,$bar) and without CO 
($X_{\rm CO}=0$).  These two spectra
are very nearly identical.   A third spectrum is computed in the unrealistic  limit where all
the carbon in the atmosphere is in the form of CO ($X_{\rm CO}=2.97 \times 10^{-4}$), which represents
the maximum possible CO 
enhancement.  As we found for NH$_3$ and H$_2$S, there is no distinctive  spectral signature
of CO at this resolution ($R=2100$).  Our extreme case represents a flux reduction of
$\sim 2\sigma$ at best.  We are unable to constrain the CO abundance with our data.
Since the $\Delta \nu=3$ band of CO is $\sim 10^3$ times weaker than the fundamental
band at 4.7$\,\mu$m,  obtaining a useful CO abundance from $H$ band spectroscopy will be a difficult
undertaking.

\subsection{$K$ band spectrum}

Of our three near infrared spectra, the $K$ band spectrum is formed 
highest in the atmosphere: $T=950$ -- 1020$\,$K  and $P \sim 3\,$bar (Fig. 6). 
This is the same level as is
probed with 4.7$\,\mu$m spectroscopy.  As in the $H$ band, the spectrum contains mainly H$_2$O
features on the blue side ($\lambda \wig< 2.1\,\mu$m) and a strong CH$_4$  band appears at
longer wavelengths.  There are also several NH$_3$ features for $\lambda < 2.05\,\mu$m
and the models predict an isolated feature of H$_2$S (Fig. 3).  

Figure 13 compares a spectrum computed for model B with [M/H]$=-0.3$ 
with the entire $K$ band spectrum. There is an excellent agreement in the structure 
of the spectrum even though the overall shape is not very well reproduced. For 
$\lambda < 2.05\,\mu$m, the model predicts strong features of NH$_3$ which we discuss in the
next section.  Beyond 2.12$\,\mu$m is a CH$_4$ band which is too weak in the
model.  The structure within the modeled band is remarkably similar to the observed spectrum,
however.
This much better agreement of the CH$_4$ band than we obtained in the $H$ band is due to
two factors. First, in this band the CH$_4$ opacity is computed from a line list, and
can therefore be computed as a function of temperature rather than at a fixed value of 300$\,$K.
Second, the lower temperature where the band is formed (Fig. 6) reduces the effect of the
incompleteness of the CH$_4$ line list above 300$\,$K.  Nevertheless, we do not consider the
CH$_4$ features here and limit our analysis to $\lambda \le 2.10\,\mu$m.

The $K$ band spectrum provides a unique opportunity:
Once the metallicity is determined by matching the depth of
the H$_2$O features between 2.05 and 2.10$\,\mu$m, the abundance of NH$_3$ can be obtained
by fitting its features below 2.05$\,\mu$m.  
The fit of the metallicity is shown in Fig. 14 for model B, which shows [M/H]=0 and $-0.5$ (top
panel) and our best fit, [M/H]=$-0.3$ (bottom panel).  Values for models A and C are given
in Table 2.  All features in this region are due
to H$_2$O and, as we found in the red spectrum and in the $J$ and $H$ bands, the oscillator 
strengths of the H$_2$O line list do not reproduce the relative strength of the observed 
features very well. 

\subsubsection{Ammonia}

In the wavelength range shown in Fig. 15, the spectrum consists of a few NH$_3$ features 
on a  background of H$_2$O absorption.
Synthetic spectra predict seven strong NH$_3$ features
in this spectrum, three of which are clearly present
(2.033, 2.037 and 2.046$\,\mu$m), one is absent (2.041$\,\mu$m) and three appear to be
missing (2.026, 2.029 and 2.031$\,\mu$m). This constitutes an ambiguous detection of NH$_3$
in Gl 229B.  The determination of the abundance of NH$_3$ from the
$K$ band spectrum is hampered by the limited accuracy of the oscillator strengths of 
the H$_2$O line list and by the incompleteness of our NH$_3$ line list for temperatures
above 300$\,$K.  The effect of the former can be seen in the trio of features at 
2.026, 2.029 and 2.031$\,\mu$m, which overlap H$_2$O absorption features. Even after removing
all NH$_3$ opacity, these features are still too strong in the calculated spectrum (top panel
of Fig. 15). The top panel of Fig. 15 as well as Fig. 13 show that for model B,
the abundance of NH$_3$ derived from the
chemical equilibrium for the adopted metallicity of [M/H]$=-0.3$ is too high.  We therefore
consider a depletion in NH$_3$ in the atmosphere of Gl 229B at the level probed by the $K$
band spectrum.  Following the approach used in fitting the $H$ band spectrum,
we express this depletion as a fraction of the chemical equilibrium abundance of NH$_3$
for the metallicity obtained independently from the amplitude of the
H$_2$O features in the $K$ band.   This fraction is applied uniformly throughout the atmosphere for the
computation of the synthetic spectrum.  Given the ambiguous presence of NH$_3$ in the
$K$ band, we have determined the optimal NH$_3$ abundance by minimizing the $\chi^2$ of the
spectral fit for $ 2.022 \le \lambda \le 2.049\,\mu$m.  This gives a 
depletion factor of $\wig<0.4$ with no NH$_3$ present being an acceptable fit.
Restricting the fit to the region where NH$_3$ features are  clearly observed ($2.032 \le
\lambda \le 2.049$) gives a similar result  but favors a finite value for the NH$_3$ depletion
of $\sim 0.2$.  The results are summarized in Table 2.
The lower panel of Figure 15 shows the model B fit 
obtained by reducing the NH$_3$ abundance throughout the atmosphere
to $\sim 25$\% of its chemical equilibrium value {\it for the adopted
metallicity}.  With this significant degree of depletion, the model reproduces
the three detected features (2.033, 2.037 and 2.046$\,\mu$m) extremely well and 
makes the 2.041$\,\mu$m feature consistent with the observations.

Because our NH$_3$ line list is incomplete at high temperatures, the opacity which we 
compute is strictly a lower limit to the actual NH$_3$ opacity at any wavelength.  If the
incompleteness is significant for the features found in the $K$ band spectrum, then the
NH$_3$ abundance is actually lower still.  It appears extremely unlikely that the errors in
the oscillator strength of the H$_2$O transitions would conspire to mimic the depletion
of NH$_3$ which we find.  For example, if we imagine that there is no NH$_3$ depletion (dotted 
curve in the top panel of Fig. 14), the residuals between the data and the fitted spectrum for
$\lambda < 2.05\,\mu$m would be
much larger than the typical mismatch which we find in H$_2$O features in all four spectra 
presented here.   We consider the depletion of NH$_3$ in the $K$ band
spectrum to be firmly established.

\subsubsection{Hydrogen sulfide}

We are not able to establish the presence of H$_2$S from our $H$ band spectrum (\S 5.3.2).
Throughout the $K$ band, the H$_2$S opacity is generally
overwhelmed by H$_2$O and CH$_4$ absorption.  However, there is a peak in the opacity of
H$_2$S which is about one order of magnitude higher than  all other opacity maxima
(Fig. 16). Our synthetic spectra indicate that this feature is strong enough to become
visible in the midst of the background of H$_2$O and CH$_4$ features.  Figure 17 shows the relevant 
portion of the $K$ band spectrum with synthetic spectra for all three 
models (A, B, and C) using the fitted metallicity (Table 2).  All three panels are
remarkably similar. The strength of the predicted feature is
well above the noise level of the data, and taken at face value, Figure 17
indicates a probable
depletion of H$_2$S by more than a factor of 2.  On the other hand, chemical equilibrium calculations
indicate that the H$_2$S abundance should be very near the elemental abundance of
sulfur, even in the presence of vertical transport and condensation \citep{fl96}.  Since there is no 
reason to expect a significant depletion of H$_2$S, the discrepancy is probably due to remaining
uncertainties in the opacities.  Our H$_2$S line list is based on an ab initio calculation
(R. Wattson, priv. comm.) which hasn't been compared to laboratory data in this part of the spectrum.
The strong feature centered at 2.1084$\,\mu$m is a blend of three strong lines from three different bands
of H$_2$S.  Possible errors in the position or strength of these lines could significantly reduce the
amplitude of the feature in our synthetic spectra.  The limitations of the background opacity of
H$_2$O and CH$_4$ may also be responsible for the observed mismatch.
Nevertheless, it is desirable  to look for this feature at a higher resolution and a higher
S/N ratio as an absence of sulfur in Gl 229B would be a most intriguing result.

\subsection{Carbon monoxide at 4.7$\,\mu$m}

Given our determination of $\Teff$ and [M/H] in terms of the surface gravity,
we can obtain the abundance of CO from the 4--5$\,\mu$m
spectra of \citet{noll} and \citet{oppen98} which is consistent with our results.  For each model
-- A, B, and C -- and using the metallicity given in Table 2, we computed synthetic spectra with various CO
abundances.  The latter is varied freely without imposing
stoichiometric constraints.  The synthetic spectra are binned to the  wavelength grid of the data and
fitted to the data by a normalization factor adjusted to minimize the $\chi^2$.  The $\chi^2$ of
the fitted spectra
shows a well-defined minimum as a function of the CO abundance, $X_{\rm CO}$ (Table 3).
The uncertainty on the CO abundance is obtained by generating synthetic data sets by adding
a gaussian distribution of the observed noise to the best fitting model spectrum.
After doing the same analysis on 1000 synthetic data sets, we obtain a (non-Gaussian)
distribution of values of $X_{\rm CO}$.  The uncertainties given in Table 3 correspond
to the 68\% confidence level.
The \citet{oppen98} spectrum gives CO abundances which are 0.1 dex higher than those obtained
from the \citet{noll} spectrum; which is well within the fitting uncertainty.  
\citet{noll} found a CO mole
fraction of 50 to 200 ppm ($-4.3 \le \log X_{\rm CO} \le -3.7$) by assuming a H$_2$O abundance 
of 300 ppm (effectively, [M/H]$=-0.6$),  which
agrees well with our result for model A which has [M/H]$=-0.5$.  Our results are also
consistent with those of \citet{griff99} who
find CO abundances of $\ge 20$ and $\ge 100$ ppm ($\log X_{\rm CO} \ge -4.7$ and $-4$) for [M/H]=$-0.6$ and 0,
respectively, using the data of \citet{noll}.

Since this part of the spectrum contains only H$_2$O and CO features, our fitting procedure
is sensitive only to the CO to H$_2$O abundance ratio.  Chemical equilibrium calculations show that
the H$_2$O abundance scales linearly with
the metallicity at the level probed with 4--5$\,\mu$m spectroscopy ({\it i.e.} all the gas phase 
oxygen is in H$_2$O).
Accordingly, the CO abundance we find scales with the metallicity of the model.    As
shown in Figure 18, the CO abundance determined from the 4.7$\,\mu$m band corresponds approximately
to the CO/CH$_4$ transition in chemical equilibrium.   The observations 
definitely exclude the very-low chemical equilibrium abundance of CO.  The fact that the extreme case
where all carbon  is CO provides an acceptable fit to the data (while we know that a good fraction of
the carbon is in CH$_4$) is due to the rather noisy spectra.

\section{Discussion}

\subsection{Surface Gravity}

It is highly desirable to restrict the surface gravity $g$ of Gl 229B to an
astrophysically useful range.
Since $L_{\rm bol}$ is known, a determination of $g$ fixes $\Teff$,
the radius, the mass, and the age of Gl 229B (Fig. 4), as well as the metallicity and the abundances of
important molecules such as CO and NH$_3$.  The large uncertainty on $g$ 
results in a large
uncertainty in the mass of Gl 229B and on the age of the system determined from cooling
tracks.  While a dynamical
determination of the mass may be possible in a decade or so \citep{golim98},
a spectroscopic determination might be obtained much sooner.

Unfortunately,  it is not possible to constrain the gravity  better than $\log g=5 \pm 0.5$
with the data and models presently available.  Our high resolution spectroscopy
does not allow us to choose between models A, B and C (Table 2) as an increase in gravity can be
compensated by an increase in metallicity to lead to an identical fit.  The gravity
sensitivity of the $K$ band synthetic spectrum models reported by \citet{ppiv} occurs
for a fixed metallicity only.

Similarly, \citet{allard96} and \citet{ppiv} report that the spectral energy distribution of Gl 229B
models is fairly sensitive to the gravity. But this is true only for a fixed metallicity.  
For the three models indicated in Fig. 4, and using the metallicity we have determined for each
(Table 2) the gravity dependence of the infrared colors is $\Delta(H-K) / \Delta (\log g)
= 0.03$ and $\Delta(J-K) / \Delta (\log g) = 0.03$ and $\Delta(K-L^\prime) / \Delta (\log g) = -0.08$
(Table 4).  This dependence is very weak in the light of the uncertainty in the photometry of Gl 229B.
More problematic is the fact that the synthetic $H-K$ disagrees with the photometry.  Furthermore,
the incomplete CH$_4$ opacities used in the spectrum calculation almost certainly result in an
inaccurate redistribution of the flux in the near infrared  opacity windows  which determines the
broad band colors.  An example of this effect on the $H$ band flux can be seen in Fig. 10.
We conclude that a  photometric determination of the gravity is not possible at present.

An alternative approach to determining the gravity of Gl 229B is through a study of the 
pressure-broadened molecular lines of its spectrum.
The spectrum of Gl 229B is formed of a forest of unresolved
molecular lines maily due to H$_2$O, CH$_4$, and NH$_3$.
Because of the limitations of the CH$_4$ and NH$_3$ opacity data bases,
a detailed study of molecular features is best performed in spectral domains where these two
molecules do not
contribute significantly to the opacity.  Spectroscopic observations with a resolution
of $\sim 50\,000$ can reveal the shape of individual H$_2$O lines in regions where
they are relatively sparse, {\it e.g.} from 2.08 to 2.105$\,\mu$m.  

\subsection{Metallicity}

Our determination of the metallicity of Gl 229B, with an uncertainty of $\pm 0.1$, is
given in Table 2 for the three gravities considered.  We find an excellent agreement
between our three independent determinations of [M/H] for each gravity and conclude that 
Gl 229B is likely to be depleted in heavy elements, {\it e.g.} oxygen.  The metallicity is near
solar at high gravity and decreases significantly for lower gravities.  In their analysis of the
``red'' spectrum, \citet{gym98} found a H$_2$O abundance between 0.3 and 0.45 of the solar value
for a $\Teff=900\,$K, $\log g=5$ model and they adopt a value of 0.25 in \citet{griff99}.

Metal depletion in Gl 229B is consistent with the analysis of the 0.985 -- 1.02$\,\mu$m FeH band
in the spectrum of the primary star Gl 229A by Schiavon, Barbuy \& Singh (1997)  who find 
[Fe/H]$=-0.2$.  The infrared colors of Gl 229A imply that it is slightly metal-rich, however 
\citep{legg92}.  The relative metallicity of the components of this binary system may have been 
affected by their formation process.  If the pair formed from the fragmentation of a collapsing cloud
(like a binary star system), the two objects should share the same composition.  If the brown
dwarf formed like a planet, from accretion within a dissipative Keplerian disk around the primary,
the selective accretion of solid phase material could lead to an {\it enrichment} in
heavy elements compared to the primary star, as is observed in the gaseous planets of the solar system.
The low mass of the primary ($\sim 0.5M_\odot$) and the large semi-major axis and eccentricity
($a \wig> 32\,$AU and $e \wig> 0.25$, \citet{golim98})
 suggests that the binary formation process, and therefore equal metallicities,
are more plausible.  A more detailed study of the metallicity of Gl 229A is desirable to better 
understand the history of this system.

\subsection{Atmospheric chemistry and molecular abundances}

The results of our analysis of the metallicity and abundances of several molecules in the
atmosphere of Gl 229B are summarized in Fig. 18.  Each panel corresponds to a different
model (see Table 2) and shows the abundances of important molecules as a function of depth
in the atmosphere based on chemical equilibrium calculations including condensation cloud formation. 
The chemistry of these abundant molecules is fairly simple.
The abundance of H$_2$O is uniformly reduced by $\sim 16$\% by silicate condensation.
Except for a small depletion for $\log T \wig< 3$ due to the condensation of Na$_2$S, 
all sulfur is found in H$_2$S. The other molecules shown are not affected by condensation.
Nitrogen is partitioned 
between N$_2$ and NH$_3$, with the latter being favored at lower temperatures and higher pressures.
NH$_3$ dominates near the surface and rapidly transforms into N$_2$ at the higher temperatures
found deeper in the atmosphere.  Deep in the atmosphere, the higher pressures cause a partial
recombination of NH$_3$ and the ratio of the NH$_3$ to N$_2$ abundances increases slowly with depth.
In a similar
fashion, all elemental carbon is found in CH$_4$ at the surface but CO starts to form at higher 
temperatures and
rapidly becomes the most abundant carbon-bearing molecule.  The formation of CO consumes
H$_2$O, as can be seen in Fig. 18.

In each panel, a dotted box indicates the CO abundance we have determined from the 4.7$\,\mu$m 
spectrum, the location of the box along the ordinate shows the level probed at this wavelength (Fig. 6).
As originally discussed by \citet{noll} and \cite{griff99}, the CO abundance is $\sim 3$ orders of
magnitude larger than the chemical equilibrium value.  Stochiometric constraints imply that this
also results in a significant reduction of the CH$_4$ abundance at the 870 -- 950$\,$K level in Gl 229B.
In model B, the equilibrium abundance of CH$_4$ at 900$\,$K is $2.97 \times 10^{-4}$, which corresponds
to the abundance of elemental carbon.  The CO abundance determined from the 4.7$\,\mu$m band 
is $1.6 \times 10^{-4}$ (with a large error bar).  Conservation of the total number of carbon
atoms then requires that the CH$_4$ abundance be $\sim 1.4 \times 10^{-4}$, a full factor of 2 below the
equilibrium abundance.  If we use the lowest CO abundance allowed by our analysis, 
a 25\% reduction of CH$_4$ relative to its equilibrium abundance at 900$\,$K results.  Depletion of CH$_4$ 
at this depth is readily accessible spectroscopically in the 1.6 and 3.3$\,\mu$m bands, and may also
affect the 2.3$\,\mu$m band if the non-equilibrium CO abundance persists at higher levels (Fig. 6).
After H$_2$O, CH$_4$ is the most important near infrared molecular absorber in Gl 229B.
Accurate modeling of the spectrum demands a careful treatment of the non-equilibrium CH$_4$ abundance.

Similarly, solid boxes show the NH$_3$ abundance determined from our $H$ and $K$ band spectra.  While the
$H$ band abundance is in excellent agreement with the equilibrium value, there is a clear
depletion of NH$_3$ in the $K$ band.  The $H$ and $K$ band abundances are marginally consistent
with each other but it appears that the NH$_3$ abundance decreases upwards through the atmosphere.

\subsection{Non-equilibrium processes}

Processes which take place faster than the time scale of key chemical reactions can drive the composition
of the mixture away from equilibrium.  The case of CO/CH$_4$ chemistry has been well-studied in the
atmosphere of Jupiter where an overabundance of CO is also observed.  Carbon monoxyde is a strongly
bound molecule and the conversion of CO to CH$_4$ through the (schematic) reaction
\begin{equation}
  {\rm CO} + 3{\rm H}_2 \rightarrow {\rm CH}_4 + {\rm H}_2{\rm O}
\end{equation}
proceeds relatively slowly, while the reverse reaction is much faster.  Vertical transport, if vigorous 
enough, can carry CO-rich gas from deeper levels upwards faster than the CO to CH$_4$ reaction can take
place.  The CO/CH$_4$ ratio at any level is fixed (``quenched'') by the condition
\begin{equation}
\tau_{\rm chem} = \tau_{\rm mix},
\end{equation}
where $\tau_{\rm chem}$ is the chemical reaction time scale  and $\tau_{\rm mix}$ the dynamical transport
time scale.  Wherever $\tau_{\rm mix } \le \tau_{\rm chem}$ in the presence of a vertical gradient
in the equilibrium abundance, non-equilibrium abundances will result.

As discussed by \citet{fl96}, \citet{noll}, \citet{griff99} (and references therein),  this naturally explains the
very high CO abundance observed at the 900$\,$K level.  In this picture, CO-rich gas would be
carried upward from $T\wig> 1400\,$K.  Convection is the most obvious form of vertical
transport in a stellar atmosphere but in Gl 229B the convection zone remains
about 3 pressure scale heights below the level where CO is observed (shaded area in Fig. 18).  
Perhaps 
convective overshooting can transport CO to the observed level.  \citet{griff99} propose
``eddy diffusion'' as a slower, yet adequate transport mechanism.  The eddy diffusion
(or mixing) time scale
is constrained by the poorly known CO abundance and the somewhat uncertain chemical pathway
between CO and CH$_4$.  From \citet{griff99}, we infer that  $\tau_{\rm mix} \wig< 1$ to 10 years
and could be much smaller.

In analogy to the CO/CH$_4$ equilibrium,
a low NH$_3$ abundance can be explained by vertical transport which can quench
the NH$_3$/N$_2$ ratio at a value found in deeper layers in the atmosphere.
As it is carried upward, N$_2$ is converted to NH$_3$ by the reaction
\begin{equation}
  {\rm N}_2 + 3{\rm H}_2 \rightarrow 2{\rm NH}_3
\end{equation}
The N$_2$ molecule is very strongly bound, however, and this reaction proceeds extremely 
slowly at low temperatures, much more slowly than reaction (2).   Thermochemical kinetic calculations
of the chemical lifetime for conversion of N$_2$ to NH$_3$ were performed as described in \citet{fl94}.
The time scale for reaction (4) along the $(P,T)$ profiles of models A, B, and C and for 
$P > 1\,$bar is given by
\begin{equation}
 \log \tau_{\rm chem} = {3.75\times 10^4 \over T} + 0.22\log g - 23.08,
\end{equation}
where $\tau_{\rm chem}$ is in year, $T$ is the temperature in K, and $g$ is the surface gravity
in cm/s$^2$.  This time scale assumes that the N$_2$ conversion occurs in the gas phase although
it could possibly be shortened by catalysis on the surface of grains.
The time scale increases very steeply with decreasing temperature.  At the level
probed in the $H$ band, $\log T = 3.1$ and $\log \tau_{\rm chem} = 7.8$ and in the $K$ band,
$\log T = 2.95$ and $\log \tau_{\rm chem} = 20.1$!  At some intermediate level (which 
depends on $g$), the time scale for the conversion of N$_2$ into NH$_3$ becomes longer than the
age of Gl 229B.  In view of the relatively very short mixing time scale inferred from the
CO abundance, it follows that the NH$_3$ abundance in the $H$ and $K$ bands is {\it entirely} determined
by non-equilibrium processes, {\it not} by reaction (4).  At depths where $\log T > 3.23$
(which corresponds to the top of the convection zone), $\tau_{\rm chem} < 1\,$yr and the reaction proceeds
fast enough to establish chemical equilibrium between NH$_3$ and N$_2$.
We therefore expect that the N$_2$/NH$_3$ ratio will be quenched at its value at $\log T \approx 3.23$,
throughout the remainder of the atmosphere.  While the time scale for eddy diffusion increases in the
upper levels of the atmosphere, the extremely long $\tau_{\rm chem}$ ensures that the NH$_3$/N$_2$ ratio
remains unchanged, regardless of how slowly the vertical mixing proceeds.  For  model B, this corresponds
to $\log X_{{\rm NH}_3}=-5.2$, in perfect agreement with the abundance found in the $H$
band (Fig. 18b).  

While the abundance of NH$_3$ determined from the $K$ band spectrum is not very precise, it is marginally
consistent with vertical mixing.  Figure 18 shows that it is more likely that the $K$ band abundance 
is smaller than found in the $H$ band, however. On the other hand, the abundances shown in Fig. 18 and 
Table 2 are depletion factors which were applied uniformly to
the chemical equilibrium abundance profile of NH$_3$, which has a large vertical gradient.  For
consistency with the mixed atmosphere picture, we have therefore redetermined 
NH$_3$ abundances using a constant abundance throughout the atmosphere and found 
$\log X_{{\rm NH}_3} \le -5.0$ and
$-4.7 \pm 0.15$ from the $K$ and $H$ band spectra, respectively (model B).  The former is in good agreement with 
our simple prediction while the $H$ band value is now rather high.  The discrepancy between the $H$ and $K$
band results thus persists in this new analysis. Changes in [M/H] within the $\pm 0.1$
uncertainty have little effect either.

We consider the possibility that this vertical gradient 
in the NH$_3$ abundance  may be caused by a different
non-equilibrium process such as the photolysis of NH$_3$ by the UV flux from the primary star.
Ammonia is a relatively fragile molecule which is easily dissociated by UV photons:  
\begin{equation}
  {\rm NH}_3 + h\nu \rightarrow {\rm NH}_2 + {\rm H}.
\end{equation}
With a photodissociation cross section of $\approx 6 \times 10^{-18}\,$cm$^2$ per molecule, optical 
depth unity for the photodissociation of NH$_3$ is reached at pressures of a few millibars in Gl 229B. 
Photodissociation of the much more
abundant H$_2$ molecules does not effectively shield NH$_3$ from incoming UV photons since the two
molecules absorb over different wavelength ranges.  Photodissociation of NH$_3$ therefore represents a
net sink of NH$_3$ which occurs at the very top of the atmosphere.  We can estimate a lower limit
to the time scale of
photodissociation of NH$_3$ by assuming that each photodissociating photon results in the destruction of
a NH$_3$ molecule.  The incident photon flux is
\begin{equation}
N_{\rm UV} = \bigg({R_\star \over d}\bigg)^2 \int_{\lambda_0}^{\lambda_1}{\lambda {\cal F}_\lambda \over hc} 
             \,d\lambda  \approx  \bigg({R_\star \over d}\bigg)^2 {\bar\lambda {\cal F}_{\bar\lambda}\over hc}
             \Delta\lambda,
\end{equation}
where photons causing dissociation of NH$_3$ are between $\lambda_0$ and $\lambda_1$, 
${\cal F}_\lambda$ is the flux at the surface of the primary star, 
$R_\star$ the radius of the primary star and $d$ the separation of the
binary system.  For NH$_3$, we have $\bar\lambda \sim 1900\,$\AA\ and $\Delta\lambda \sim 300\,$\AA\ (Moses,
priv. comm.). The primary star has a dM1 spectral type, with 
${\cal F}_{\bar\lambda} \sim 10^7\,$erg$\,$cm$^{-2}$s$^{-1}$cm$^{-1}$ and
$R_\star \sim 3.6 \times 10^{10}\,$cm \citep{legg96}.  The binary separation is $d \wig> 44\,$AU
\citep{golim98}.  This results in $N_{\rm UV} \wig< 8.3 \times 10^3\,$cm$^{-2}$s$^{-1}$.
Photodissociation will affect significantly the NH$_3$ abundance when
\begin{equation}
  \tau_{\rm phot} = {\sigma \over N_{\rm UV}} \wig< \tau_{\rm mix},
\end{equation}
where $\sigma$ is the column density of NH$_3$.  Because the incident flux of UV photons is fairly low, this
condition is satisfied only at pressures below a few microbars, {\it i.e.} very high in the
atmosphere.
In the region of interest, photodissociation destroys a very small fraction of NH$_3$  during one 
mixing time scale and therefore has little effect on the abundance of NH$_3$.  Photolysis of NH$_3$
cannot explain the relatively low NH$_3$ abundance we find in the $K$ band spectrum.

We believe that the difference between the $H$ and $K$ band determinations of the NH$_3$ abundance arise
from the limitations of the NH$_3$ opacity data used for the calculation of the synthetic spectra.
As discussed above, the incompleteness of the NH$_3$ line list for $T > 300\,$K results in upper limits
for the NH$_3$ abundances obtained by fitting the data.  Since this effect increases with
temperature, we expect that the abundance determined from the $H$ band spectrum is overestimated relative 
to the one obtained from the $K$ band spectrum, which is what we observe.  Until NH$_3$ opacities
become available for $T\sim 800$ -- 1200$\,$K, we will not be able to quantify this effect.

Figures 6 and 18 show that ammonia offers a third window of opportunity for a 
determination of its abundance.
The region between 8.3 and 14.4$\,\mu$m is rich in strong NH$_3$ features, the
two strongest being at 10.35 and 10.75$\,\mu$m (Fig. 19).  This spectral region
probes a higher level in the atmosphere ($T \sim 500 - 800\,$K) where
the hot bands of NH$_3$ which are missing from opacity data bases
are less problematic than in the $H$ and $K$ bands.  In this spectral region,
the spectrum is very sensitive to the NH$_3$/H$_2$O ratio, especially for NH$_3$ abundances
below 25\% of the equilibrium value  (Fig. 19). This would
allow a good determination of the degree of NH$_3$ depletion in the upper levels
of the atmosphere.  We anticipate that 
10$\,\mu$m spectroscopy should reveal a NH$_3$ abundance of $\wig< 10$\% of its equilibrium value.

\section{Conclusion}

With the availability of extensive photometric, astrometric, and spectroscopic data, our
picture of the atmosphere of Gl 229B is gradually becoming more exotic and more complex.
The initial discovery of CH$_4$ in its spectrum set it appart and has
prompted the creation of a new spectral class, the T dwarfs.  H$_2$O, CO,
Cs I, and K I have also been detected.  There is good evidence that the rapid decrease of the flux
at visible wavelengths is caused by unprecedently broad lines of atomic alkali metals \citep{liebert00}.
The presence of condensates may also play a role in shaping the spectrum of Gl 229B. 

It is unfortunate that the surface gravity of Gl 229B remains poorly constrained.  We have not
been able to further restrict the allowed range with our new 
$J$, $H$, and $K$ spectroscopy.  As a result, all our results are expressed as a function of
gravity.  This is the most significant obstacle to further progress in elucidating the
astrophysics of this T dwarf.  The surface gravity can probably be determined from the study of 
the pressure broadened shape of molecular lines.

We have found good evidence for the presence of NH$_3$ in the spectrum of Gl 229B, which
was expected from chemical equilibrium calculations.  We have been able to determine its
abundance at two different levels in the atmosphere, and we find a significant 
deviation from chemical equilibrium.  A similar situation has been found with CO previously
\citep{noll} and this abundance pattern can be explained by vertical mixing in the atmosphere.
The extent of the convection zone is not sufficient to account for the abundances we find
and the mixing may  be due to overshooting or to less efficient eddy diffusion.
We find that NH$_3$ photolysis is not important in shaping the spectrum of Gl 229B.
Because NH$_3$ can be observed in three different bands corresponding to three distinct
depths in the atmosphere, an accurate determination of its abundance in each band provides
information on the time scale of mixing as a function of depth.  This is an unusual and
powerful diagnostic tool which can provide valuable clues for modeling the 
vertical distributions of possible condensates.  In principle, any absorber with a large 
abundance gradient
through the visible part of the atmosphere can be used to infer the details of the mixing
process.  Among detected and abundant molecules, only CO and NH$_3$ satisfy this criterion.
Chemical equilibrium calculations with rainout of condensates \citep{lodders99,bms99}  show that
we can expect significant vertical gradients in the abundances of atomic  K, Rb, Cs, and Na
as they become bound in molecules (KCl, RbCl, CsCl and Na$_2$S, respectively) in the cooler, upper
reaches of the atmosphere.  Cesium and potassium have been detected in the spectrum of Gl 229B,
and resonance doublets of K I and Na I appear to shape the visible spectrum. 
However, the chemical timescales for alkali metals are so short
that they should always remain in thermodynamic equilibrium (Lodders 1999). 
Therefore, they cannot serve as probes of vertical mixing in Gl 229B.

Further progress in understanding the atmosphere of Gl 229B requires better opacities for
CH$_4$ and NH$_3$, and, to a lesser extent, of H$_2$O.   A more accurate determination of the
CO abundance from 4 -- 5$\,\mu$m spectroscopy is very desirable and will require higher
signal-to-noise spectroscopy than is currently available.  Similarly, 10$\,\mu$m spectroscopy to
determine the NH$_3$ abundance for $P \wig < 1\,$bar, while difficult, is important.

The issue of vertical mixing and departures from chemical equilibrium gains importance when we
consider that the observed departure of CO from chemical equilibrium implies a significantly
reduced CH$_4$ abundance, by conservation of the abundance of elemental carbon.  
Similarly, our results imply that NH$_3$ absorption in the
10$\,\mu$m region is reduced.  Because CH$_4$ is a significant absorber in the near infrared, as is  
NH$_3$ in the 10$\,\mu$m range, departures from equilibrium must
be taken into account when accurate modeling of the atmosphere and spectrum of Gl 229B is
desired.  This new level of complexity compounds the exoticism and the challenges posed by T dwarfs.

The astrophysics of Gl 229B is far richer than has been originally anticipated.  
Gl 229B is currently the only T dwarf known to be in a binary system.  There is no evidence that
the illumination from the primary star has a significant effect on the state of
its atmosphere and Gl 229B is most likely typical of isolated T dwarfs.  It remains
the brightest and by far the best studied of the seven T dwarfs currently known, but  the list
should expand to several dozens during the next 2 -- 3 years \citep{burgasser, strauss}.
The existing body of work
on Gl 229B points to the most rewarding observations to conduct on T dwarfs.  The possibility
of studying trends in the physics of T dwarf atmospheres as a function of effective temperature 
is a fascinating prospect.  

\acknowledgements 

We thank T. Guillot for
sharing programs which were most useful to our analysis, J. Moses for invaluable
information regarding the photolysis of NH$_3$ in giant planets, and K. Noll and 
B. Oppenheimer for sharing their data. We are grateful to the staff at the United Kingdom 
Infrared Telescope, which is operated by the Joint Astronomy Center Hawaii on behalf of
the UK Particle Physics and Astronomy Research Council.
This work was supported in part by NSF grants AST-9318970 and AST-962487 and
NASA grants NAG5-4988 and NAG5-4970. Work by B. Fegley and K. Lodders is supported by
grant NAG5-6366 from the NASA Planetary Atmospheres Program.

\newpage

\figcaption{
$J$ band spectrum.  The spectral resolution is $R=2400$ and the noise level at the maximum flux level
is indicated by the error bar.  Nearly all features
in this spectrum are caused by H$_2$O absorption, except at short wavelengths where the CH$_4$
opacity becomes significant, and for two strong K I lines. The flux calibration is approximate.}

\figcaption{
Same as Fig. 1 for the $H$ band.  The features seen in the spectrum are caused by H$_2$O
for $\lambda \wig< 1.59\,\mu$m.  At longer wavelengths, the rapid decrease in flux 
corresponds to the blue side of a strong CH$_4$ absorption band. Both NH$_3$ and H$_2$S have
significant opacity in this spectral range but show no distinctive feature at this resolution.
The flux calibration is approximate.}

\figcaption{
Same as Fig. 1 for the $K$ band.  The features seen in the spectrum are caused by H$_2$O
for $\lambda < 2.11\,\mu$m.  At longer wavelengths, the rapid decrease in flux 
corresponds to the blue side of a strong CH$_4$ absorption band. Several strong NH$_3$ features are
seen shortward of 2.06$\,\mu$m.  Models predict a single H$_2$S absorption feature at
$2.1084\,\mu$m. The flux calibration is approximate.}

\figcaption{
Evolution of solar metallicity brown dwarfs and giant planets in the effective temperature -- gravity
plane.  The
heavy solid lines are cooling tracks for objects with masses of 0.075, 0.07, 0.06,
0.05, 0.04, 0.03, 0.02, and 0.01$\,M_\odot$, from top to
bottom, respectively (Burrows et al. 1997).  Evolution proceeds from right to left 
and isochrones (dotted lines) are labeled with the age in Gyr.  
The band crossing the center of the figure is the locus of all models
with the luminosity of Gl 229 B $(L_{\rm bol}=6.2 \pm 0.55 \times 10^{-6}\,L_\odot)$.
Filled symbols correspond to the models listed in Table 2 and used for the present 
analysis.}

\figcaption{
Temperature-pressure profiles of the three atmosphere models used in this analysis.  Labels
correspond to the models shown in Fig. 5 and Table 2.}

\figcaption{
The temperature and pressure at the photosphere $(\tau_\lambda=2/3)$ as  a function
of wavelength for model B ($\Teff=940\,$K, $\log g=5$).  The temperature plotted (left axis)
is equivalent to a brightness temperature.  Horizontal bars indicate the wavelength range
of NH$_3$ (solid lines) and CO (dotted lines) bands.  The short vertical bar 
indicates a H$_2$S absorption feature at $2.1084\,\mu$m (see \S 5.4.2).}

\figcaption{A spectrum computed with model B, [M/H]$=-0.3$, and dust opacity (thin line)
is compared to the observed spectrum of Oppenheimer et al. (1998) (heavy line).  Top panel:
Entire fitted region.  Bottom panel: details of the H$_2$O absorption band at a
resolution $R=2250$.}

\figcaption{Top panel: Effect of the metallicity on the $J$ band spectrum.  Data are
compared to two model B spectra with [M/H]$=-0.4$ and 0.  Bottom panel: The best fit for
model B 
is obtained for [M/H]$=-0.2$.  See text.  The synthetic spectra are normalized to the
observations at 1.2472$\,\mu$m and are plotted at the same resolution ($R=2400$).}

\figcaption{Effect of the NH$_3$ abundance on the $H$ band spectrum. The spectrum is computed
for model B with [M/H]$=-0.3$ with the chemical equilibrium abundance of NH$_3$ (dotted
line) and without NH$_3$ opacity (solid line).  The dotted curve represents our best fit
for this value of [M/H]. The synthetic spectra are normalized to the
observations at 1.5744$\,\mu$m and plotted at the same resolution ($R=2100$).} 

\figcaption{Top panel: Effect of the metallicity on the $H$ band spectrum.  Data are
compared to two model B spectra with [M/H]$=-0.4$ and 0.  Bottom panel: The best fit for
model B is obtained for [M/H]$=-0.3$.  In all cases, the NH$_3$ abundance is determined from the
chemical equilibrium. See text.  See Fig. 9 caption for additional details.}

\figcaption{Effect of the H$_2$S abundance on the $H$ band spectrum. The spectra are computed
for model B with [M/H]$=-0.3$ with the chemical equilibrium abundance of H$_2$S 
and without H$_2$S opacity.  See Fig. 9 caption for additional details.}

\figcaption{Carbon monoxide absorption band in the $H$ band. The spectra are computed
for model B with [M/H]$=0$ with the chemical equilibrium abundance of CO (solid line) 
and with the CO abundance equal to the elemental carbon abundance (dotted line).   A spectrum
computed with the CO abundance set to zero overlaps the  equilibrium CO curve.
See Fig. 9 caption for additional details.}

\figcaption{Synthetic spectrum for model B with [M/H]$=-0.3$ compared to the observed 
$K$ band spectrum.  The synthetic spectrum is normalized to the
data at 2.0513$\,\mu$m and plotted at the same resolution ($R=2800$).}

\figcaption{Top panel: Effect of the metallicity on the $K$ band spectrum.  Data are
compared to two model B spectra with [M/H]$=-0.5$ and 0.  Bottom panel: The best fit for
model B is obtained for [M/H]$=-0.3$.  All features in this figure are caused by H$_2$O.
See Fig. 13 caption for additional details.}

\figcaption{Portion of the $K$ band spectrum which shows NH$_3$ features.  The data are
 shown by the heavy solid line, with a representative error bar.  Ammonia
 features are indicated by tick marks.  Top panel: Synthetic spectrum
 computed from model B with [M/H]$=-0.3$
 (dotted line) and after removing all NH$_3$ opacity (thin solid line).
 Bottom panel: Same as above but the equilibrium NH$_3$ abundance has been uniformly
 reduced by a factor of 0.25 throughout the atmosphere. 
 See Fig. 13 caption for additional details.}

\figcaption{Opacity of H$_2$S in the 2.11$\,\mu$m region for $T=800\,$K and $P=1\,$bar.
The spectral resolution is $R=4.8 \times 10^5$.
Synthetic spectra predict that the strong feature at 2.1084$\,\mu$m should be observable.}

\figcaption{Hydrogen sulfide feature predicted at 2.1084$\,\mu$m.
The observed spectrum is shown by a heavy solid line.
The thin lines in each panel show spectra computed for the indicated  model
(A, B, and C).  Each spectrum is computed with the best-fitting metallicity for that model
(Table 2).  In each panel, the upper curve is calculated without H$_2$S opacity,
and the lower curve is computed with the chemical equilibrium abundance of H$_2$S for
the chosen metallicity.  For model B (middle panel), the effect of reducing the equilibrium
H$_2$S abundance by factors of 0.5 and 0.25, respectively, is also shown. All synthetic
spectra are normalized to the data at 2.1062$\,\mu$m and the resolution is $R=2800$.}

\figcaption{Equilibrium chemistry in the atmosphere of Gl 229B.  The curves show the
abundance of the important molecular species as a function of depth for each
of the three gravities, computed with the metallicity we have determined for each (Table 2).
Depth increases toward the right along the ordinate
axis and is indicated by both temperature and  pressure scales.  The extent of the convection
zone is shown by the shaded region.  Horizontal bars at the top show the regions probed by
spectroscopic observations of various molecular bands.  The top row of solid bars shows
three regions where NH$_3$ is detectable: 10$\,\mu$m region and the $K$ and $H$ bands, from
left to right, respectively. Dotted bars show the CO bands at 4.7$\,\mu$m and in the $H$ band,
respectively.  The vertical tick shows the depth where the 2.1084$\,\mu$m feature of H$_2$S 
is formed. The dotted box shows the CO abundance determined from the at 4.7$\,\mu$m  band
spectra of Noll et al. (1997) and Oppenheimer et al. (1998).
Our determinations of the NH$_3$ abundance in the $K$ an $H$ bands are shown by vertical boxes,
the height of the box indicating the estimated uncertainty.  For the $K$ band measurements,
an arrow indicates that the NH$_3$ abundance could be close to zero.  The labeling of the
panels corresponds to the entries in Table 2.}

\figcaption{Ammonia features in the $10\,\mu$m region.  The spectra are computed with
         model B and [M/H]$=-0.4$.  Each curve corresponds to a different depletion
         of NH$_3$ relative to its equilibrium abundance, the abundance of
         all other compounds
         being kept the same.  The equilibrium abundance of NH$_3$ has been multiplied
         by 0, 0.25, 0.5, and 1, from top to bottom, respectively. All spectra are
         normalized at 10$\,\mu$m and the resolution is $R=200$.}
\newpage

\tablecolumns{6}
\begin{deluxetable}{cccccc}
\tablewidth{0pt}
\tablecaption{Observing Log}
\tablehead{
\colhead{UT Date} & \colhead{wavelength} & \colhead{Resolving} &
\colhead{Integ. Time} & \colhead{mean airmass} & \colhead{mean airmass} \\
\colhead{} & \colhead{($\mu$m)} & \colhead{Power} & \colhead{(minutes)} &
\colhead{Gl~229B} & \colhead{BS~1849}
}

\startdata

19980125 & 2.10-2.18 & 3200 & 80 & 1.37 & 1.39 \\
19980125 & 2.02-2.10 & 3100 & 93 & 1.40 & 1.35 \\
19980126 & 1.53-1.61 & 2350 & 53 & 1.46 & 1.39 \\
19980126 & 1.25-1.30 & 3000 & 53 & 1.33 & 1.32 \\
19980126 & 1.20-1.25 & 2900 & 53 & 1.41 & 1.40 \\
19980127 & 2.02-2.10 & 3100 & 93 & 1.36 & 1.32 \\

\enddata
\end{deluxetable}

\tablecolumns{11}
\begin{deluxetable}{ccrccccccrr}
\tablewidth{0pt}
\tablecaption{Optimal parameters}
\tablehead{
\colhead{Model} & \colhead{$\log  g$}  &  \colhead{$T_{\rm eff}$}  & \multicolumn{4}{c}{[M/H]\tablenotemark{a}}
 & \colhead{} &
  \multicolumn{3}{c}{NH$_3$/$\{{\rm NH}_3\}_{\rm eq}$} \\ 
\cline{4-7}  \cline{9-11} \\
\colhead{}  & \colhead{(cgs)}  & \colhead{(K)} &   \colhead{$0.92\,\mu$m}  & \colhead{$J$}  & \colhead{$H$}  & \colhead{$K$} & &  \colhead{$H$\tablenotemark{b}} & \colhead{$H$\tablenotemark{c}} &  \colhead{$K$}}  
\startdata
 A & 4.5 &  870 & $-0.5$ & $ -0.5$ & $-0.3$ & $-0.5$ & &   0\phd\phn\phn   & 0.25 &  $\wig<0.40 $ \\
   &     &      &        &         & $-0.4$ &        & & 0.25  & 0.5\phn &                  \\
   &     &      &        &         & $-0.5^*$ &        & & 0.5\phn  & 1\phd\phn\phn &       \\
 B & 5.0 &  940 & $-0.3$ & $ -0.2$ & $-0.1$ & $-0.3$ & & 0\phd\phn\phn & $\wig<0.25$ & $\wig<0.40$  \\
   &     &      &        &         & $-0.2$ &        & & 0.25 - 0.5\phn & 0.5\phn &         \\
   &     &      &        &         & $-0.3^*$ &        & & $>0.5$\phs\phs & 1\phd\phn\phn &   \\
 C & 5.5 & 1030 & $-0.2$ &\phs 0.1 & \phs 0.0  & $-0.1$ & &  0.5\phn  & 0.5\phn & $\wig<0.40$ \\
   &     &      &        &         & $-0.1^*$ &        & &  1\phd\phn\phn  & 1\phd\phn\phn &    \\
\enddata
\tablenotetext{a}{An asterisk (*) indicates the adopted metallicity.}
\tablenotetext{b}{H$_2$S abundance set to 0.}
\tablenotetext{c}{H$_2$S abundance from chemical equilibrium.}
\end{deluxetable}

\tablecolumns{5}
\begin{deluxetable}{ccrcc}
\tablewidth{0pt}
\tablecaption{Abundance of CO from the 4.7$\,\mu{\rm m}$ band}
\tablehead{
\colhead{Model} & \colhead{$\log  g$}  &  \colhead{$T_{\rm eff}$}  & {[M/H]} & $\log X_{CO}$ \\
\colhead{}  & \colhead{(cgs)}  & \colhead{(K)} &  \colhead{}  & \colhead{} } 
\startdata
 A & 4.5 &  870 & $-0.5$ & $-4.0 \pm 0.25$ \\
 B & 5.0 &  940 & $-0.3$ & $-3.8 \pm 0.3$ \\
 C & 5.5 & 1030 & $-0.1$ & $-3.5 \pm 0.3$ \\
\enddata
\end{deluxetable}

\tablecolumns{7}
\begin{deluxetable}{ccrcccc}
\tablewidth{0pt}
\tablecaption{Near Infrared Colors of GL 229B}
\tablehead{
\colhead{Model} & \colhead{$\log  g$}  &  \colhead{$T_{\rm eff}$}  & {[M/H]} & $J-K$ & $H-K$ & $K-L^\prime$}
\startdata
 A & 4.5 &  870 & $-0.5$ & $-0.11$ &  $-0.27$ & 2.25 \\
 B & 5.0 &  940 & $-0.3$ & $-0.14$ &  $-0.30$ & 2.26 \\
 C & 5.5 & 1030 & $-0.1$ & $-0.08$ &  $-0.24$ & 2.17 \\
 \multicolumn{4}{c}{Leggett et al. 1999}& $-0.10 \pm 0.07$ &  $-0.07 \pm 0.07$ & $2.24 \pm 0.11$ \\
\enddata
\end{deluxetable}

\clearpage

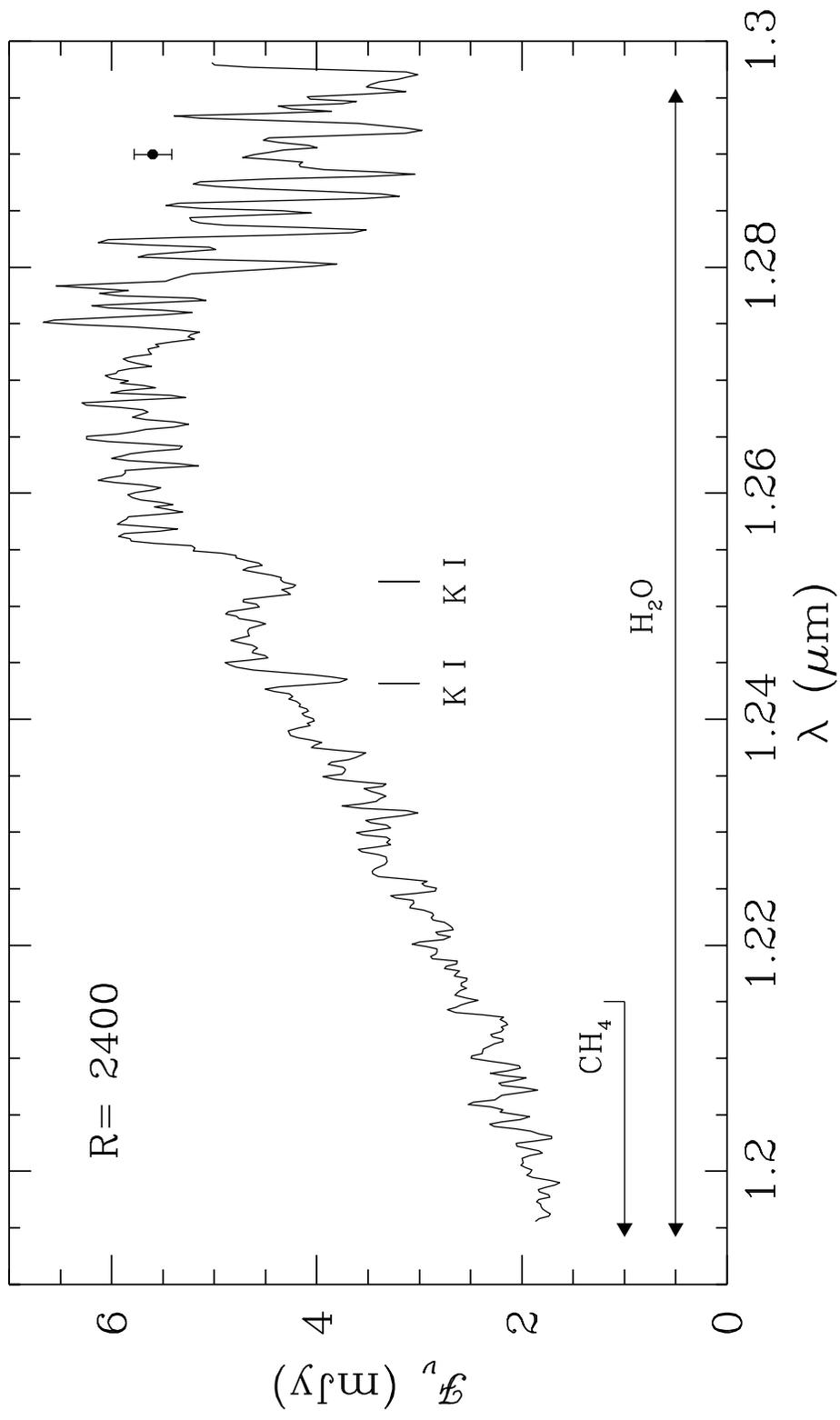
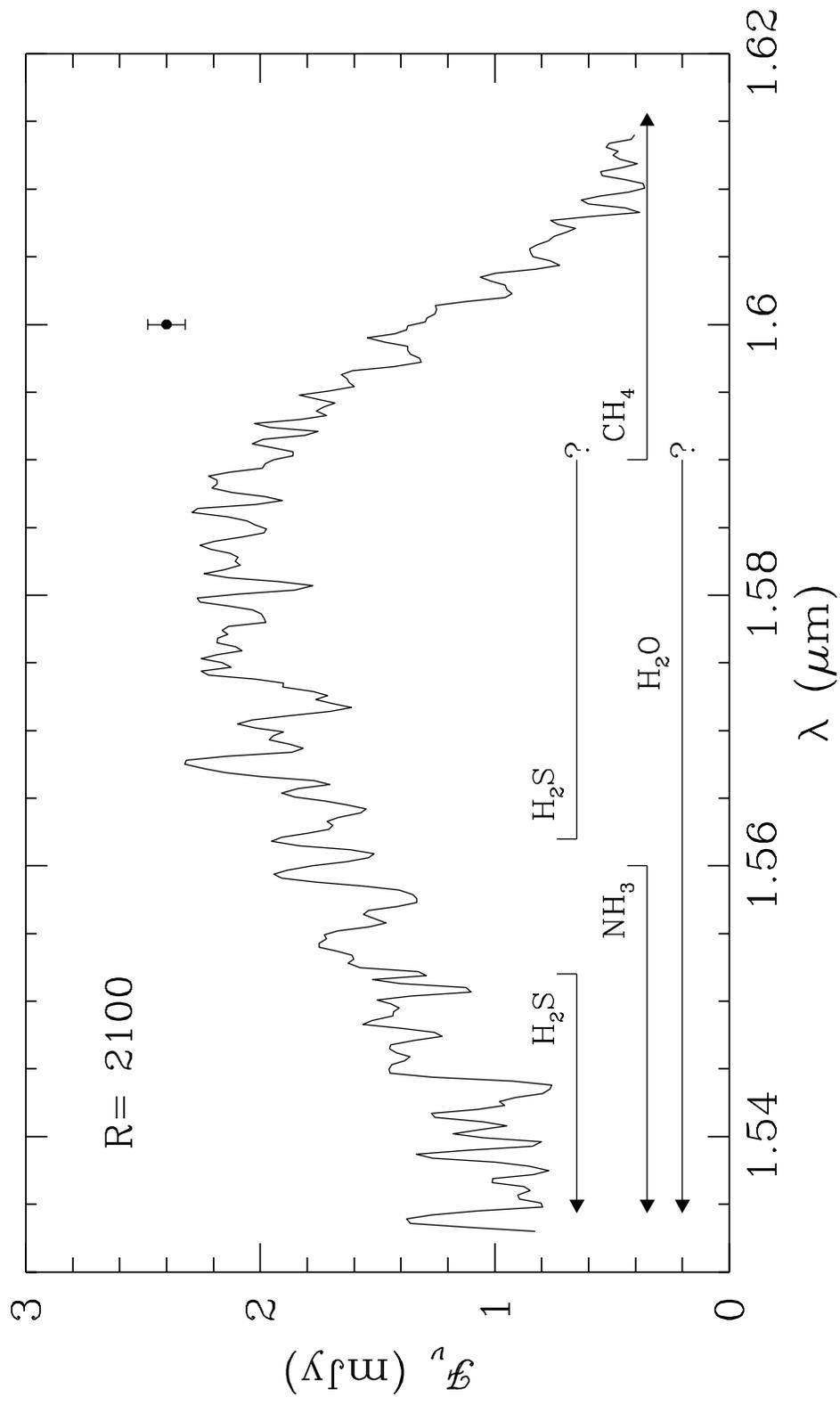
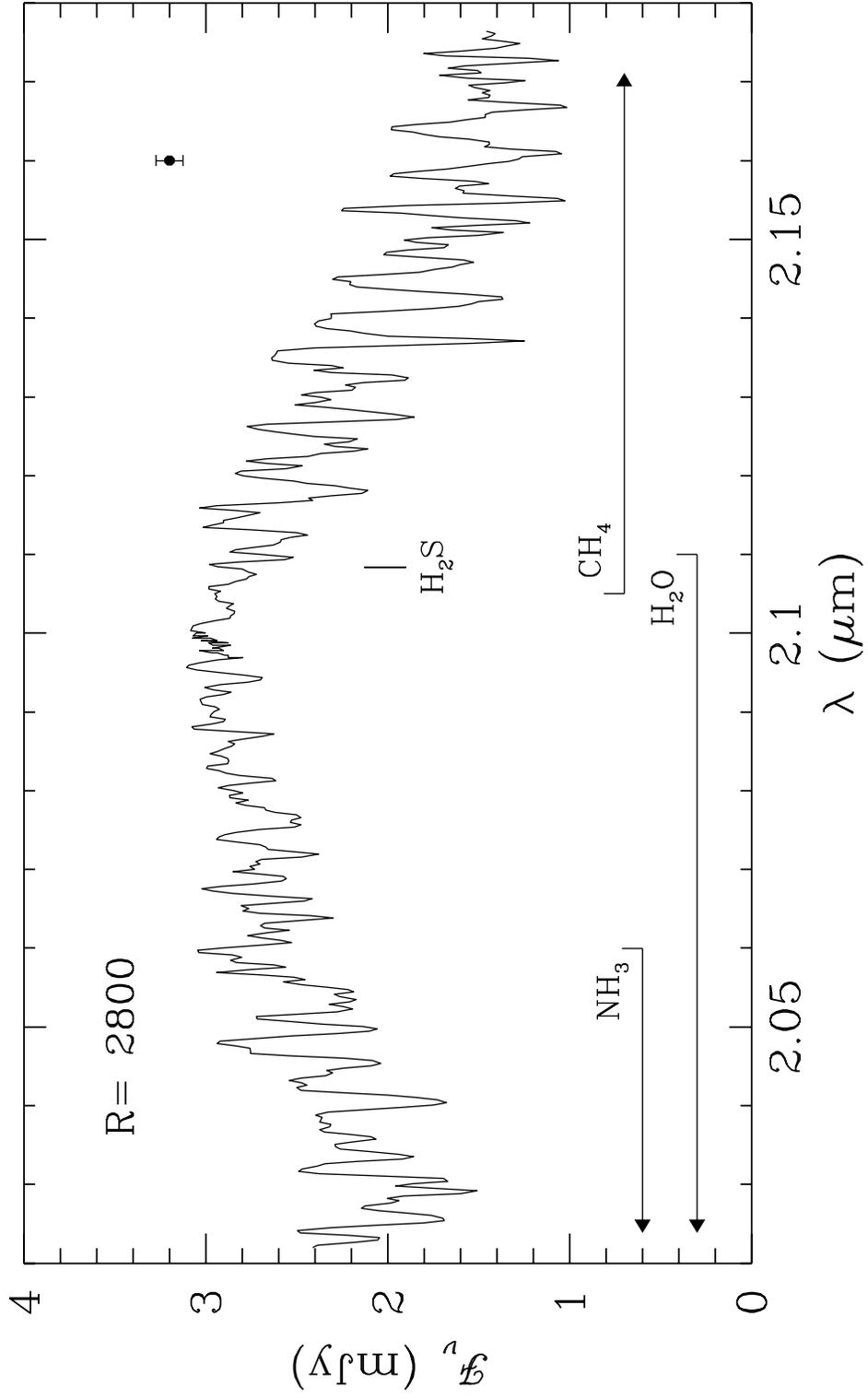
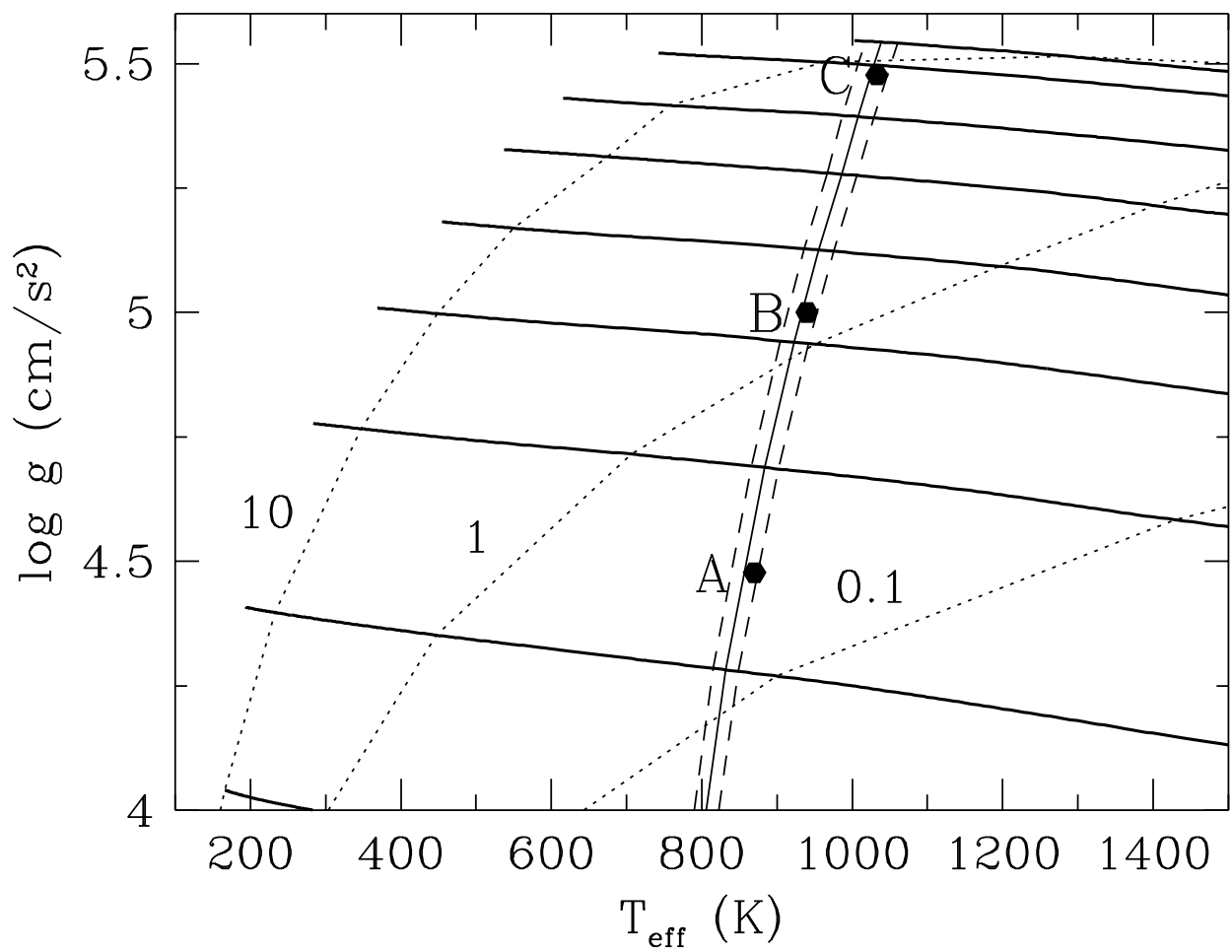
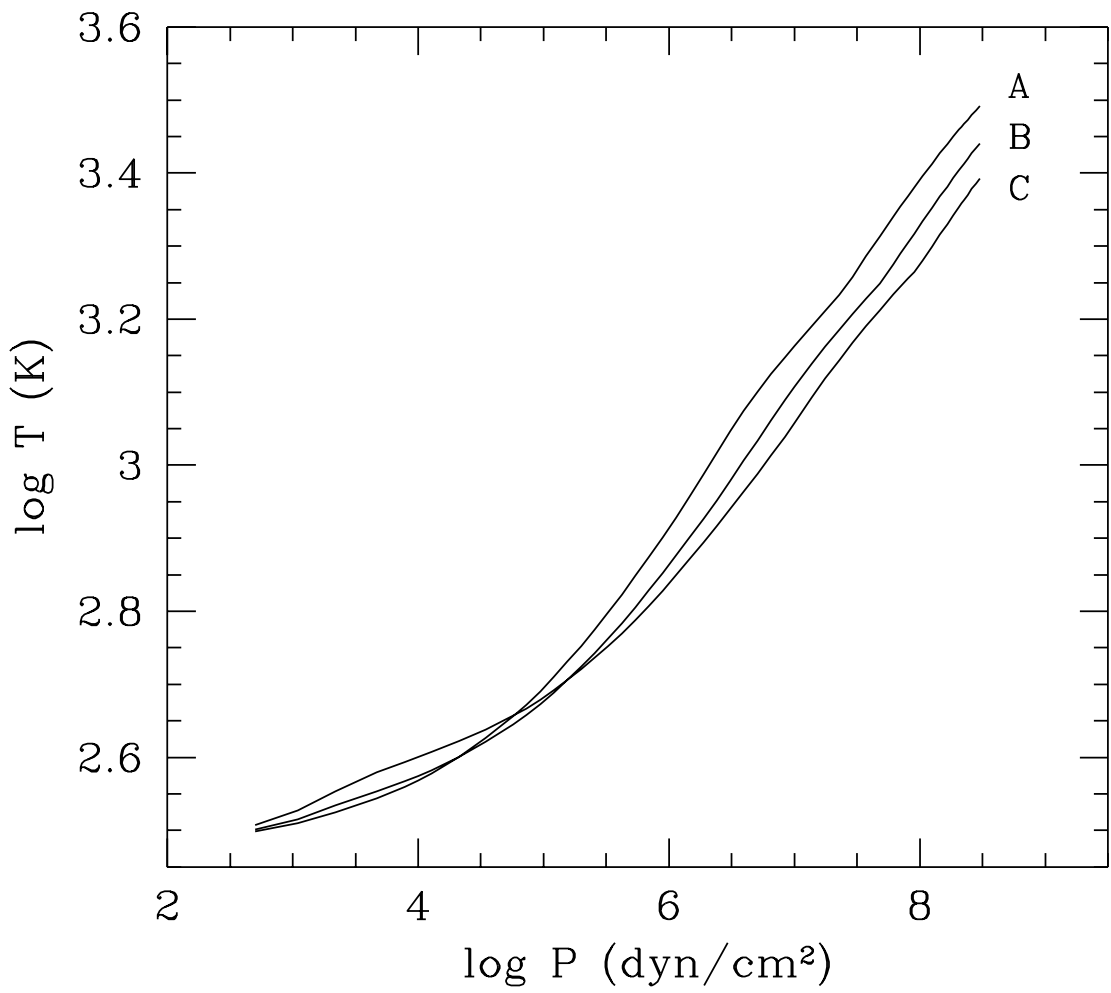
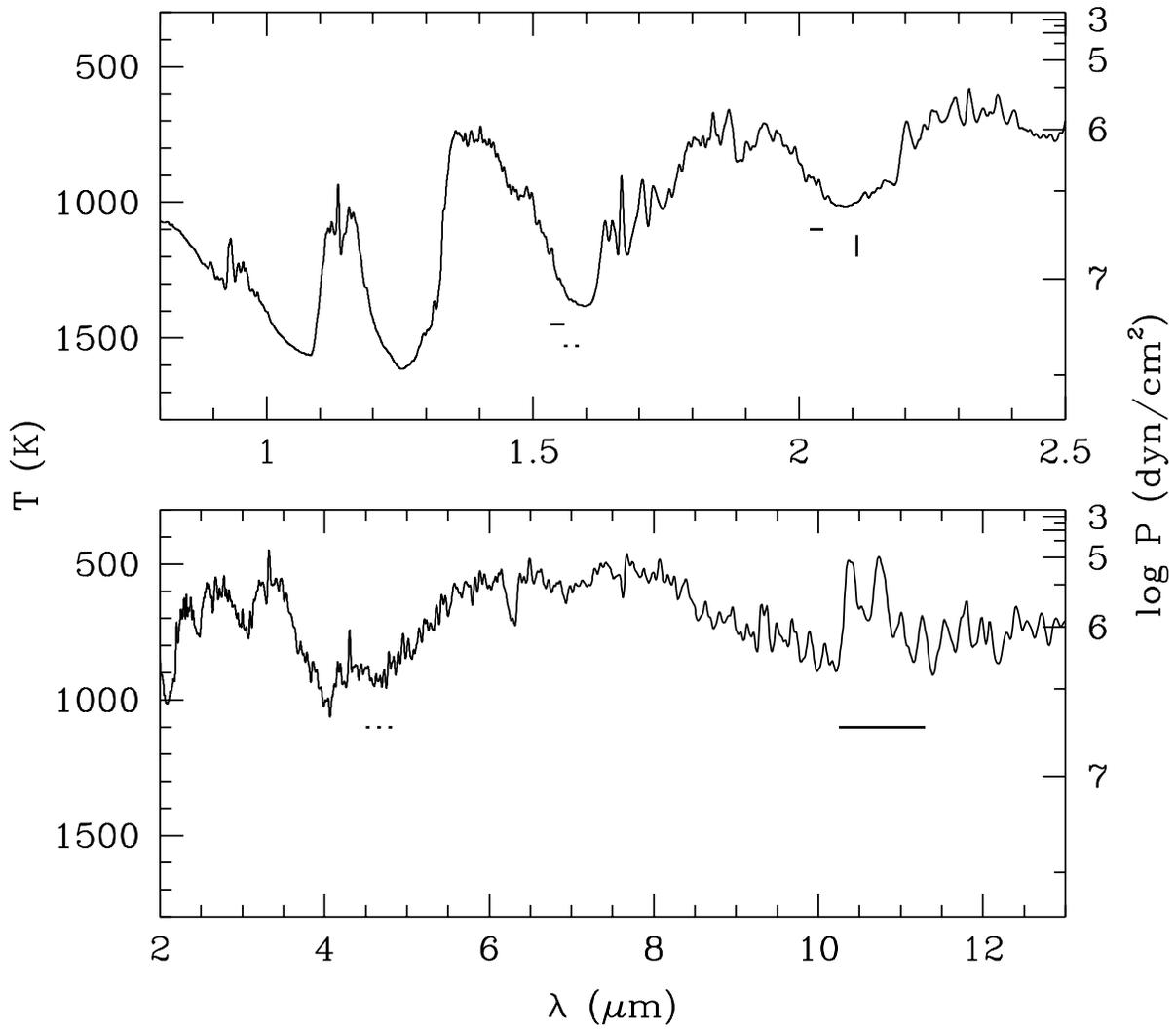
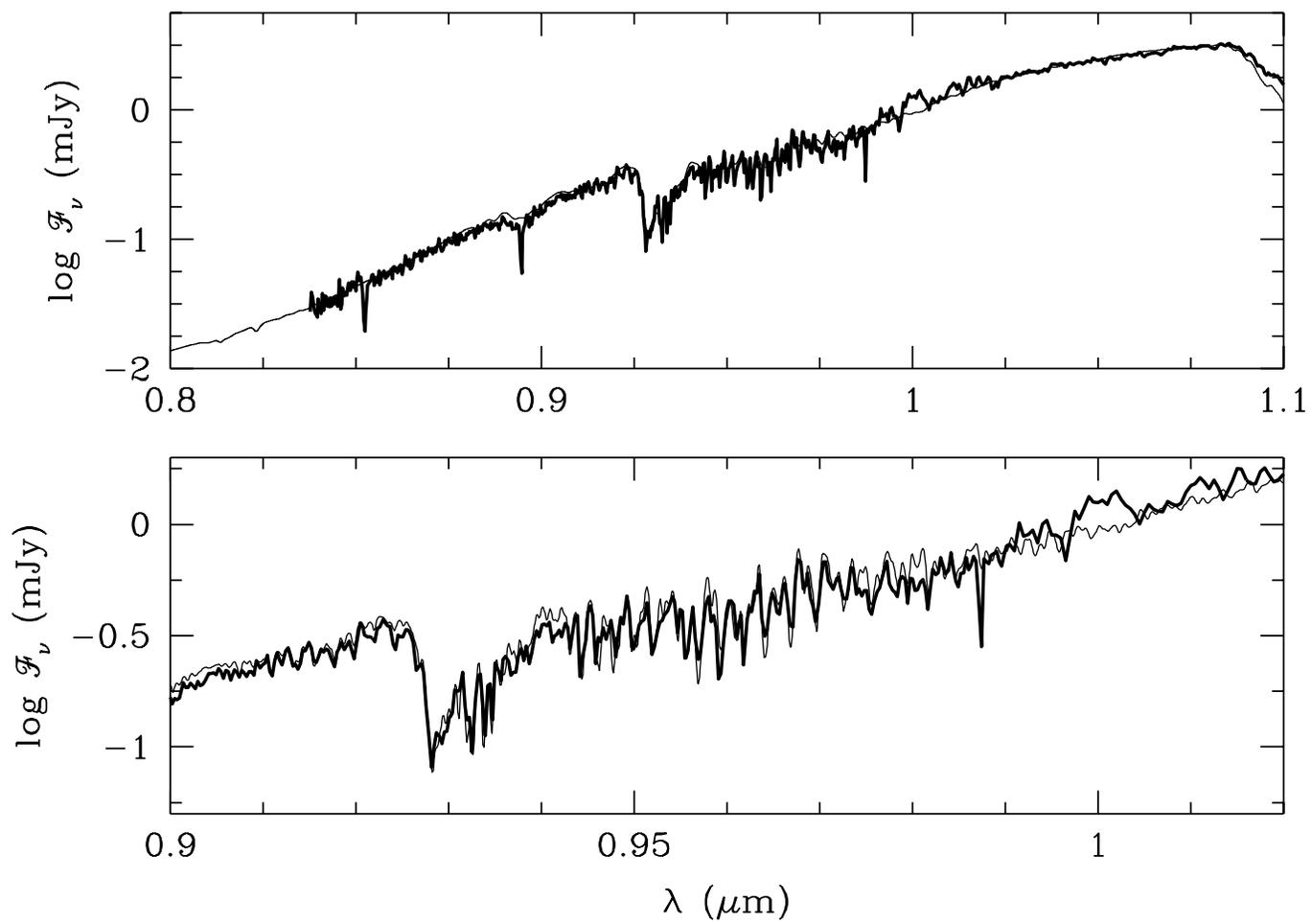
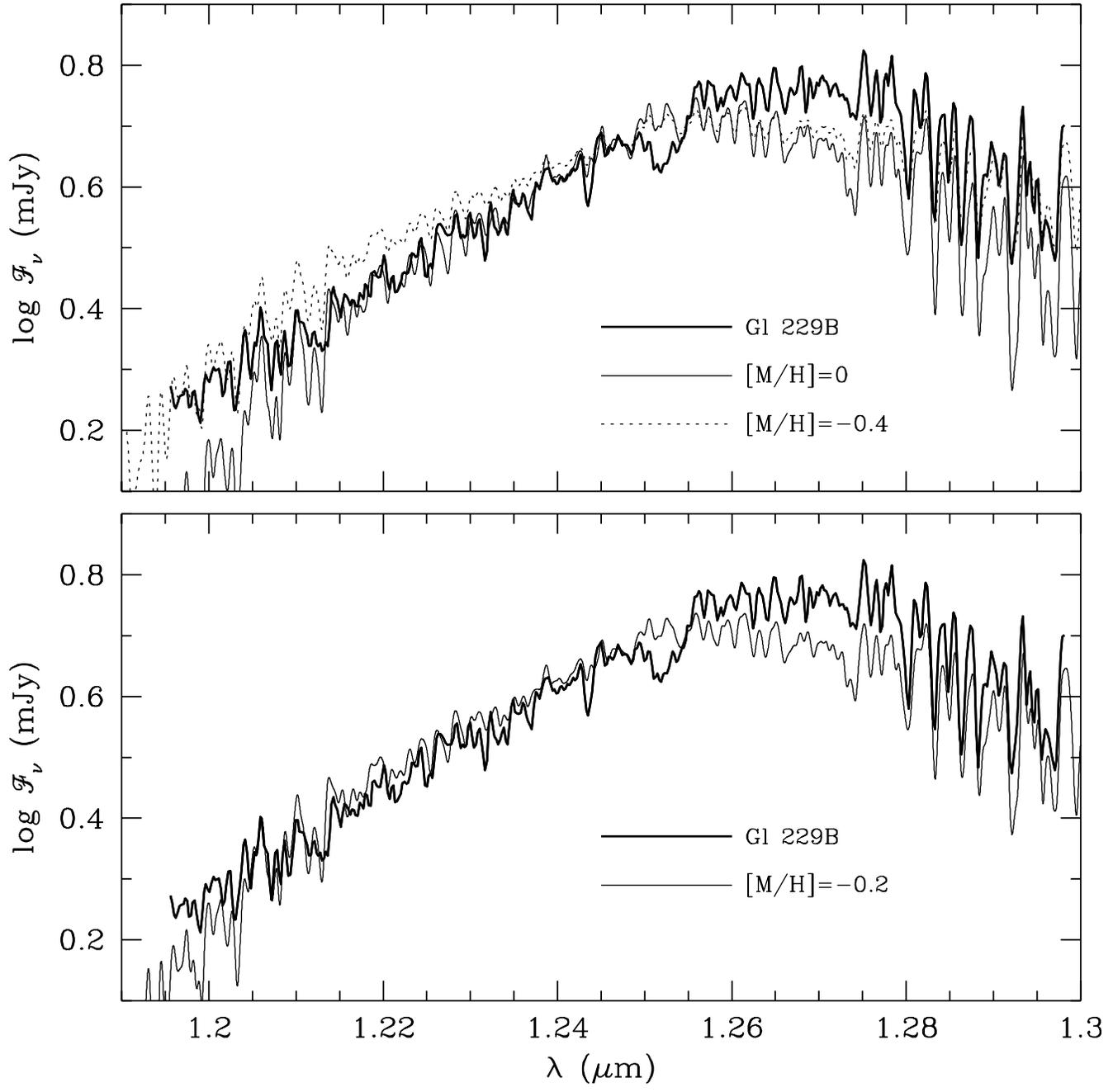
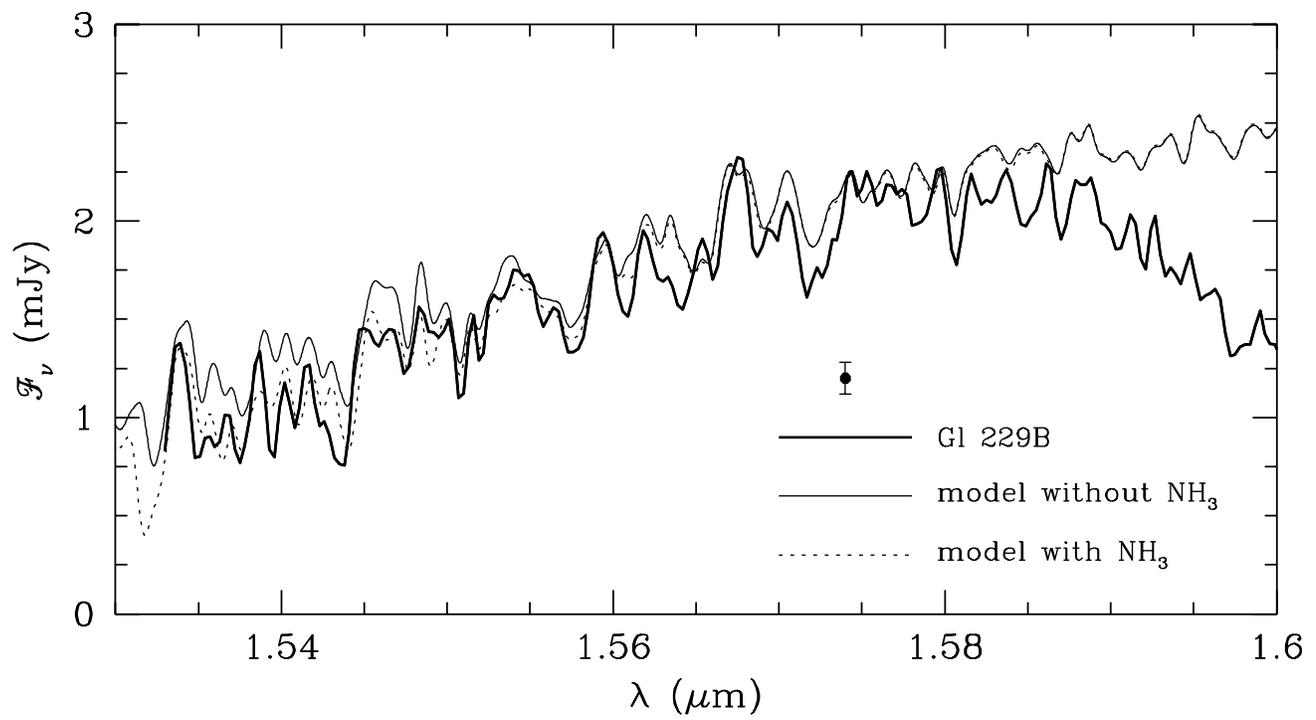
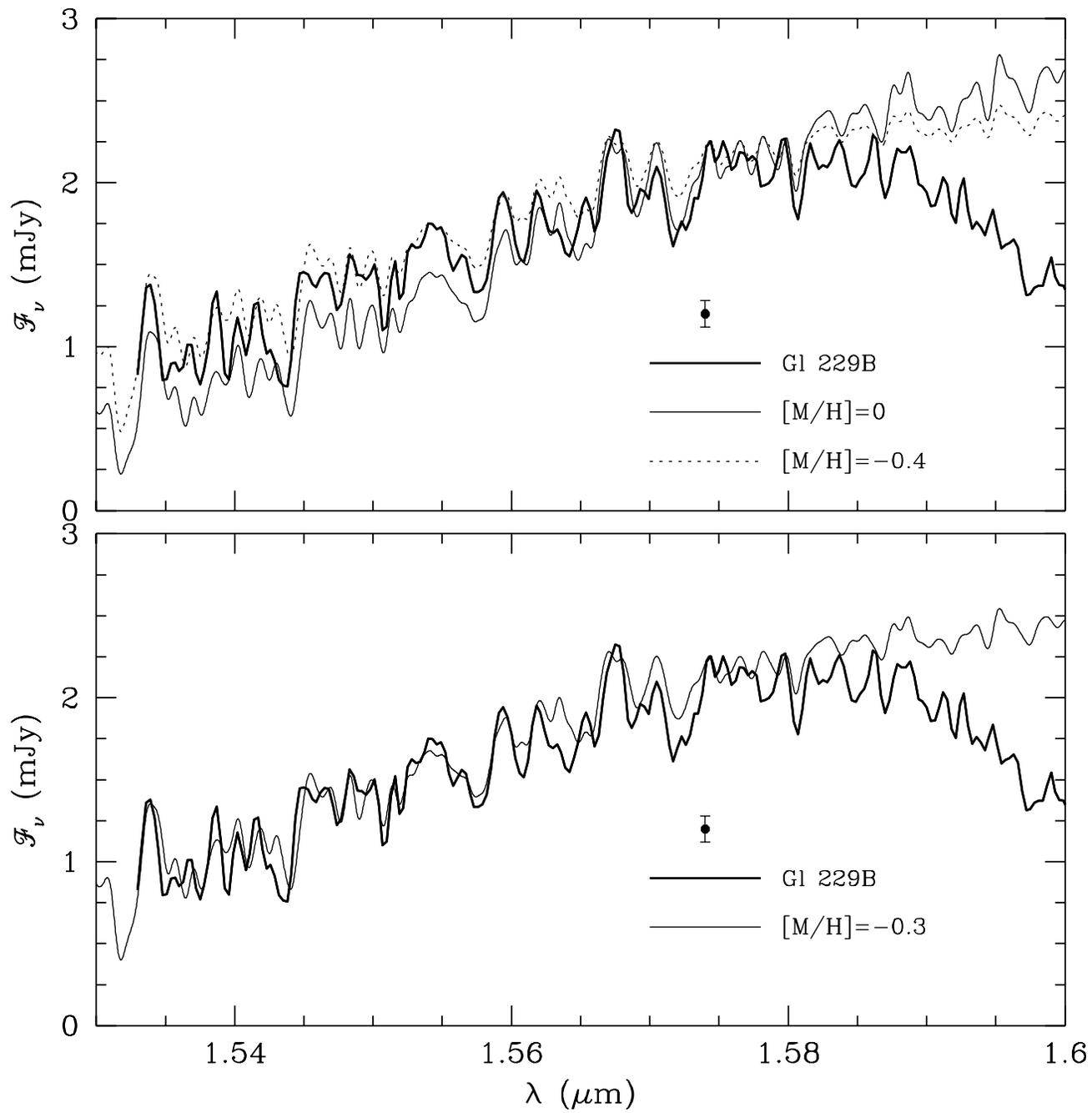
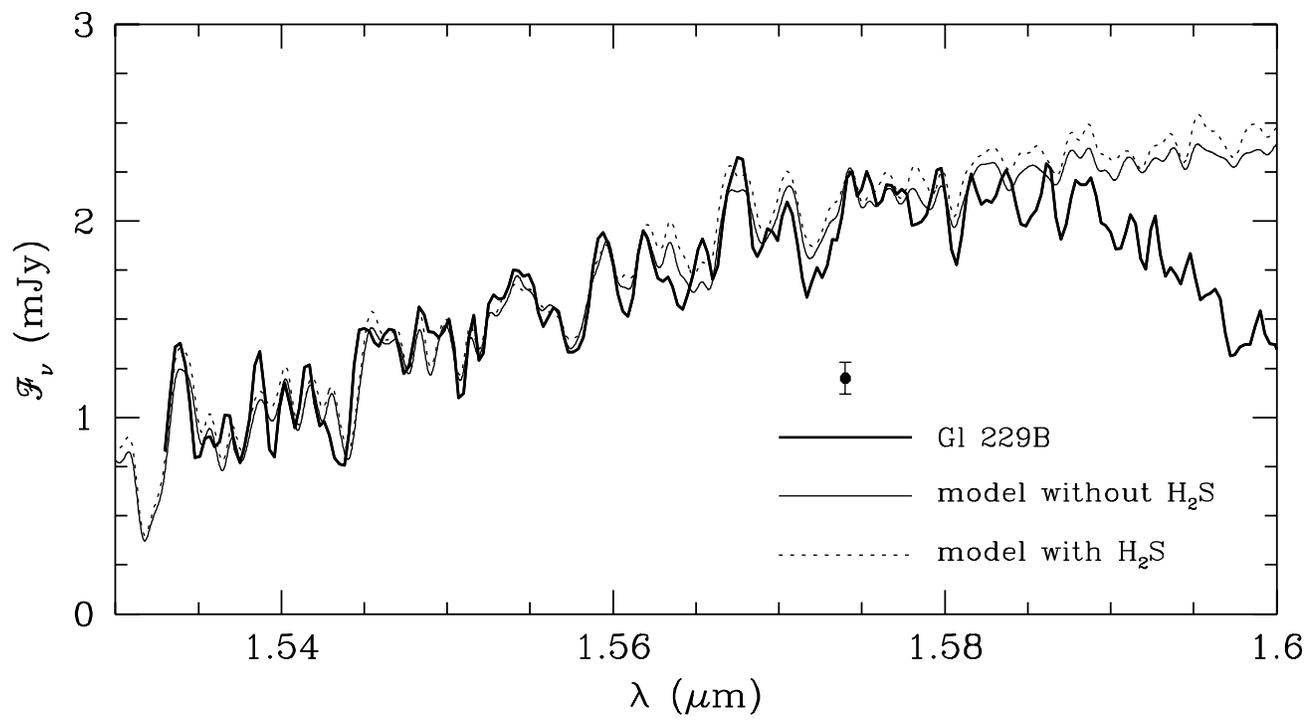
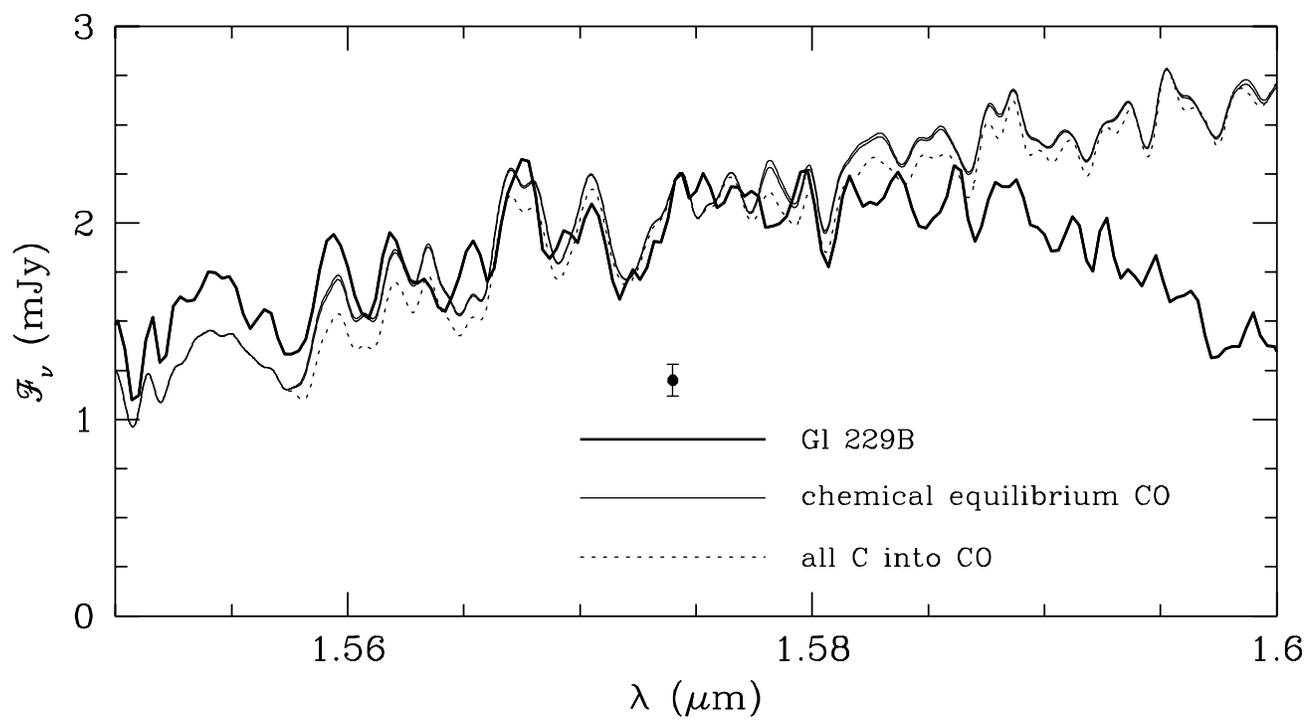
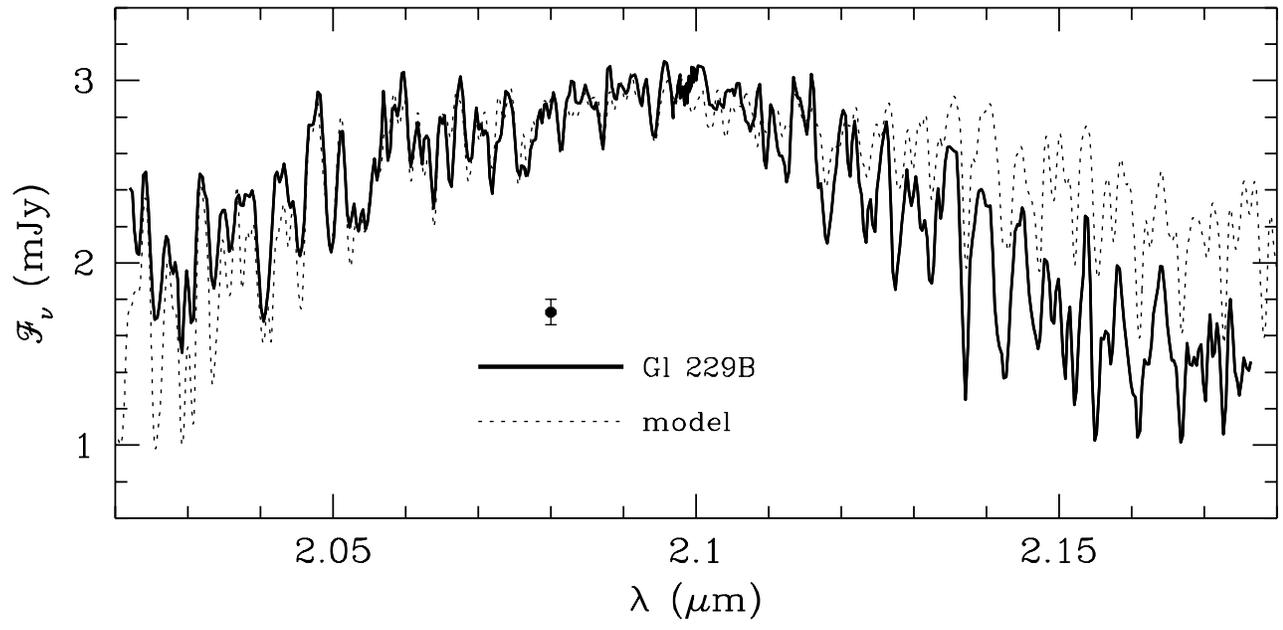
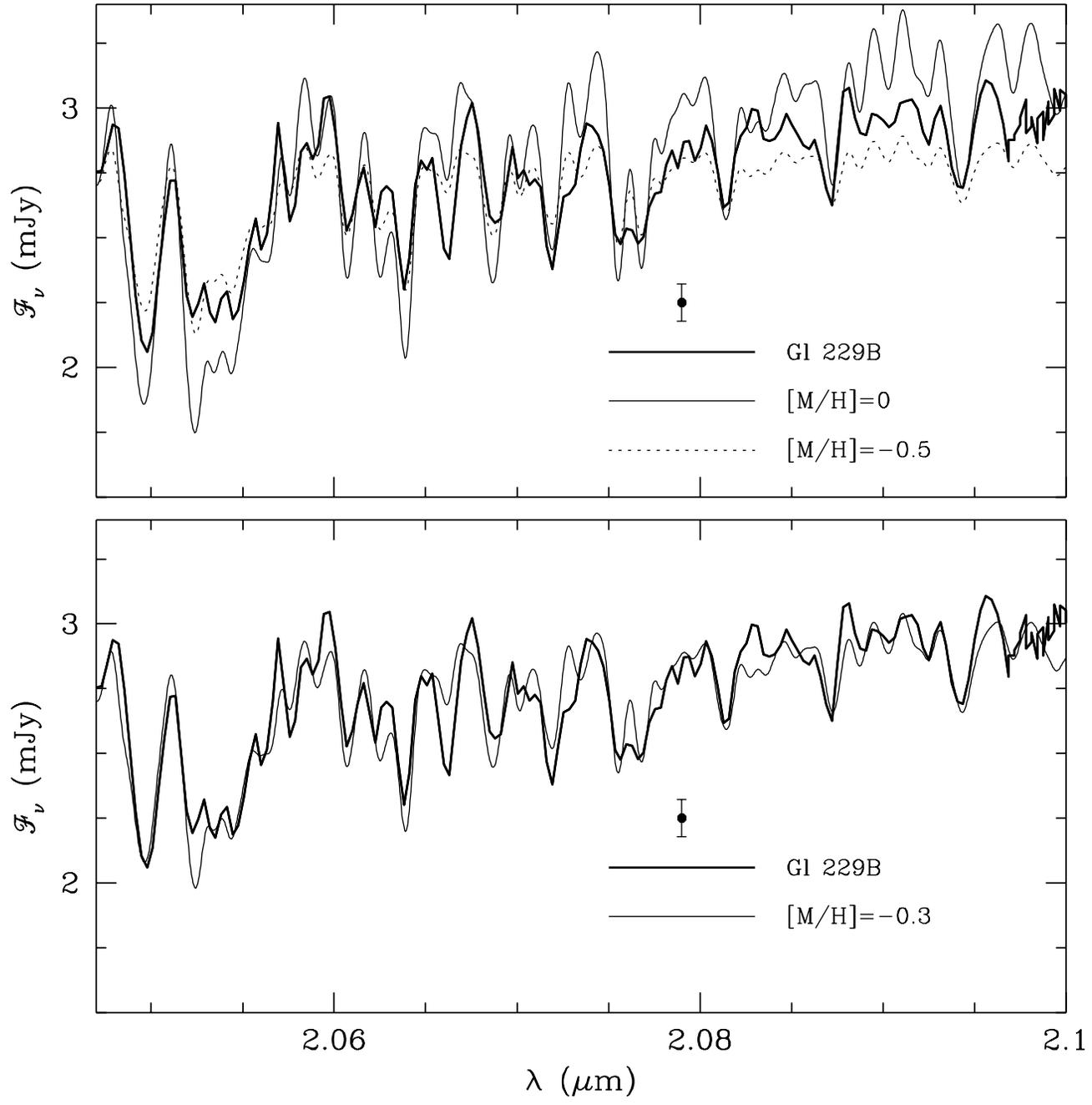
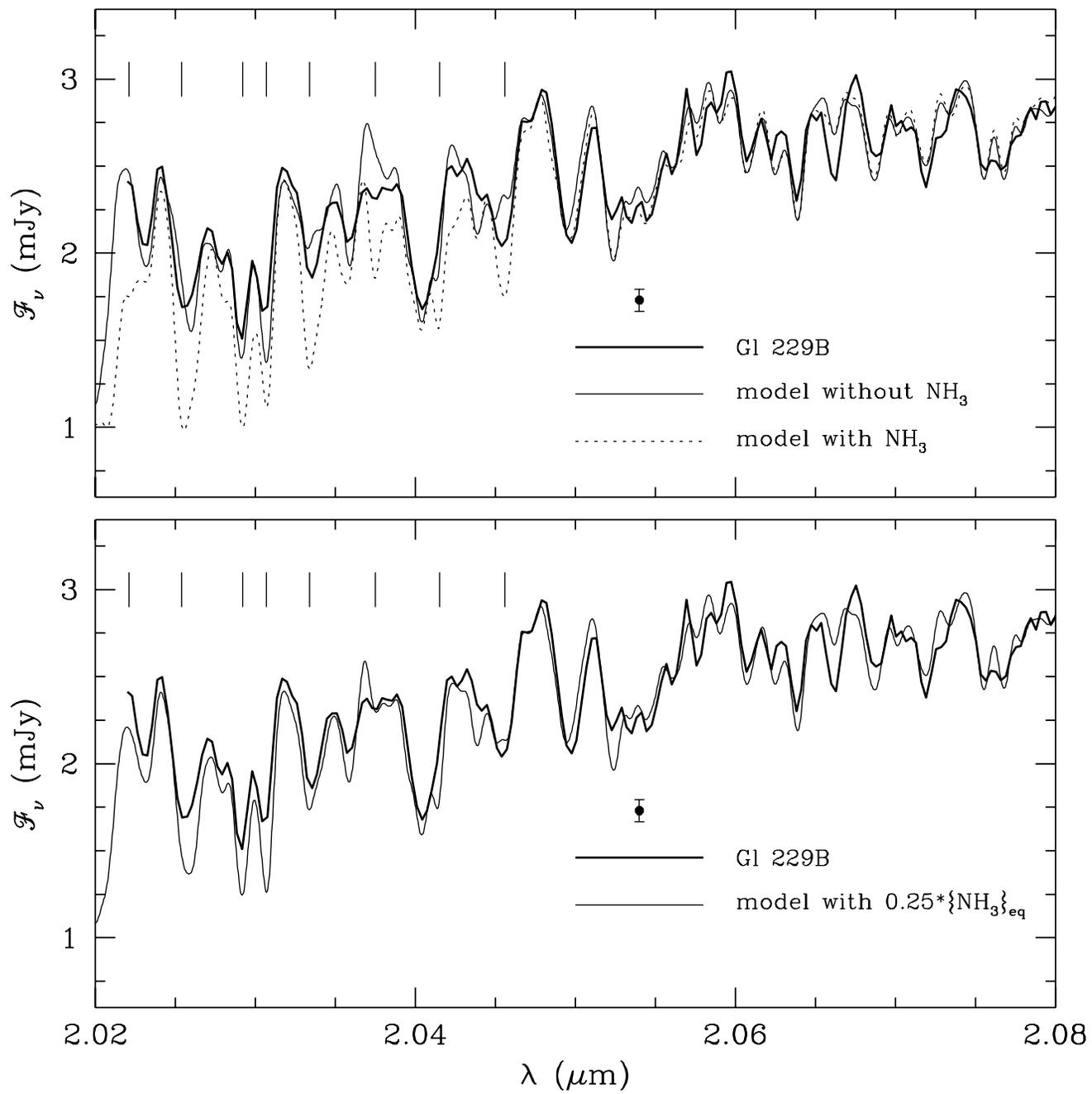
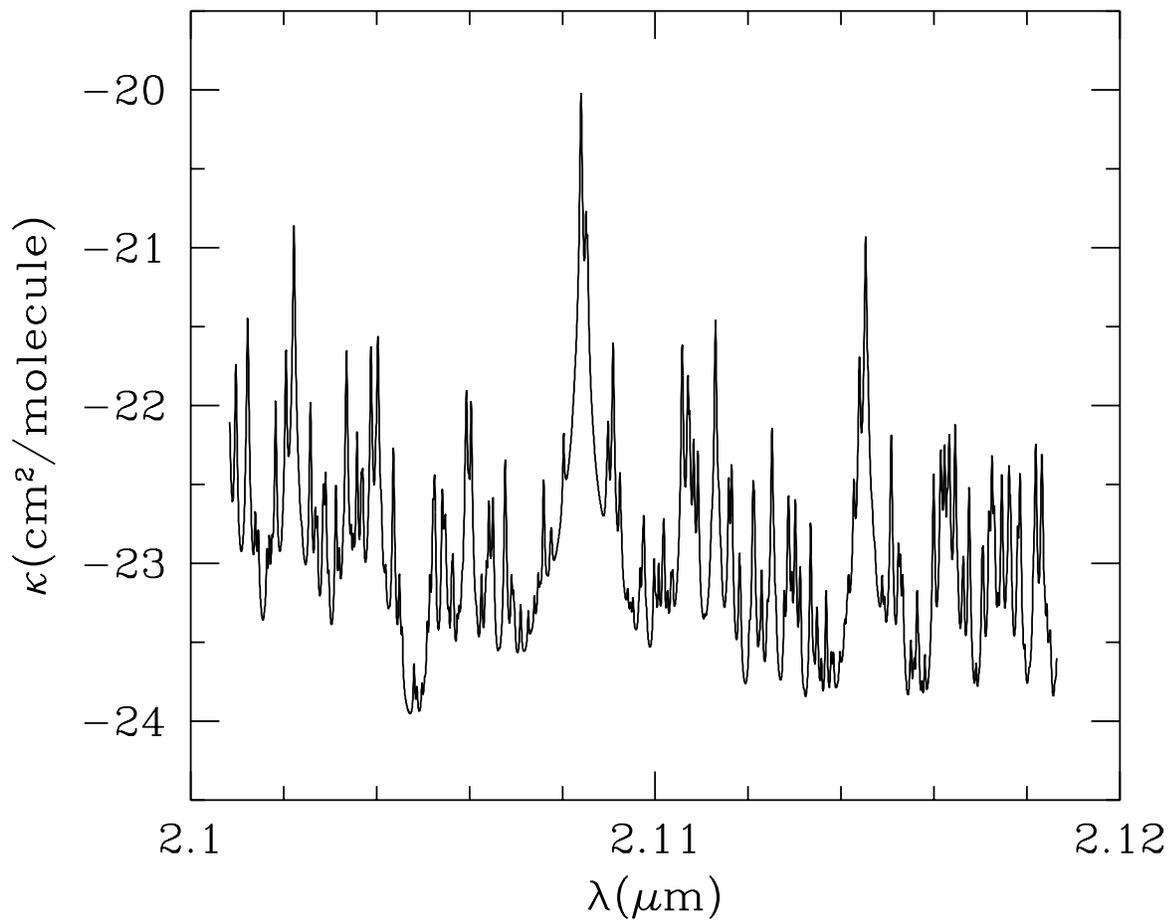
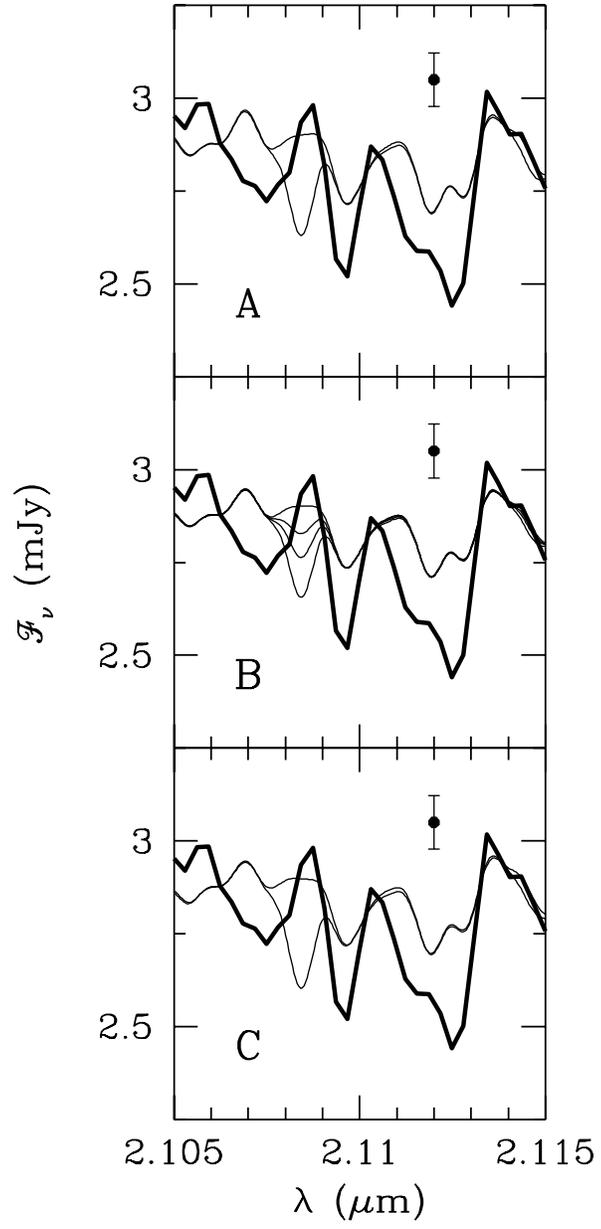
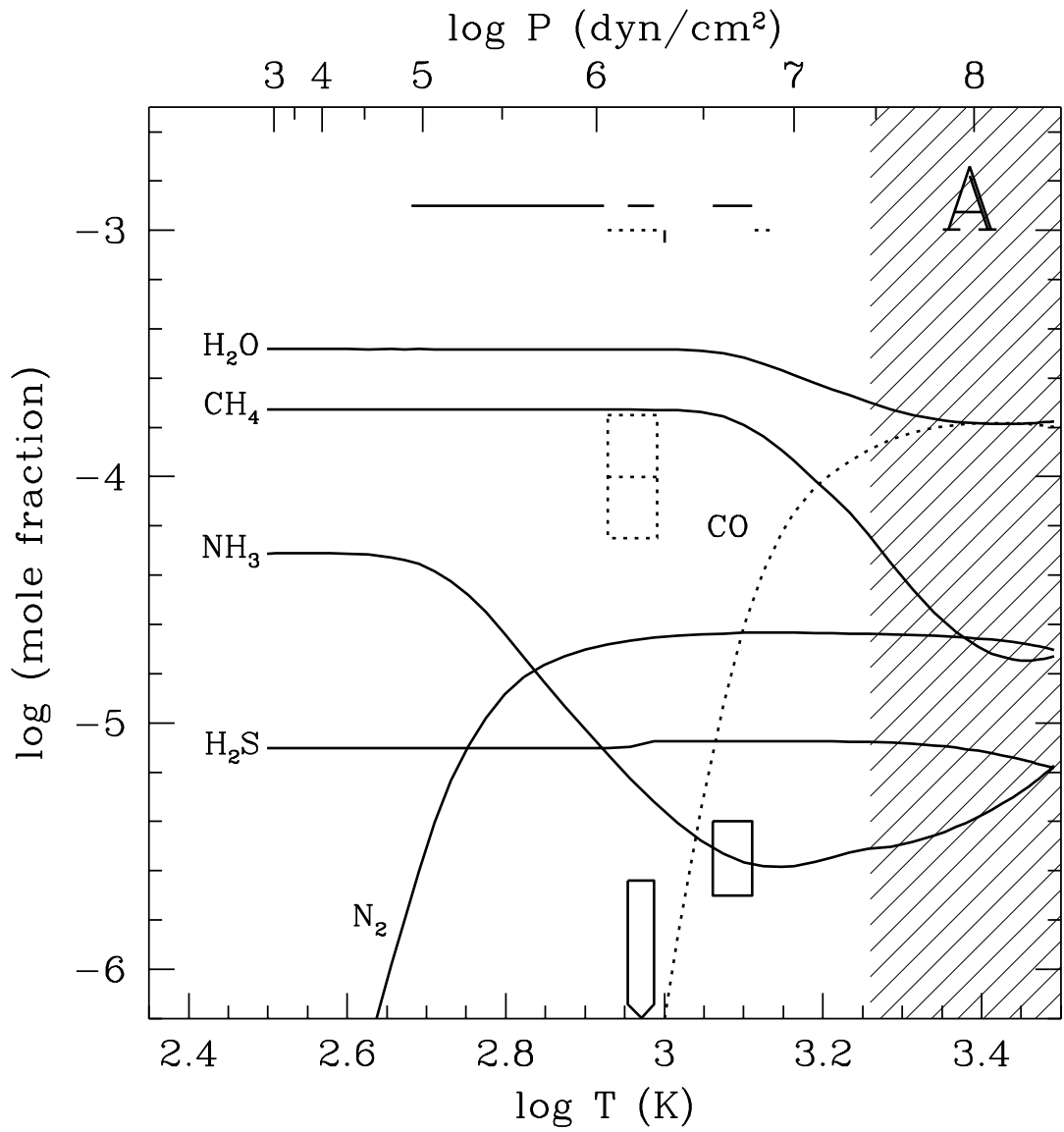
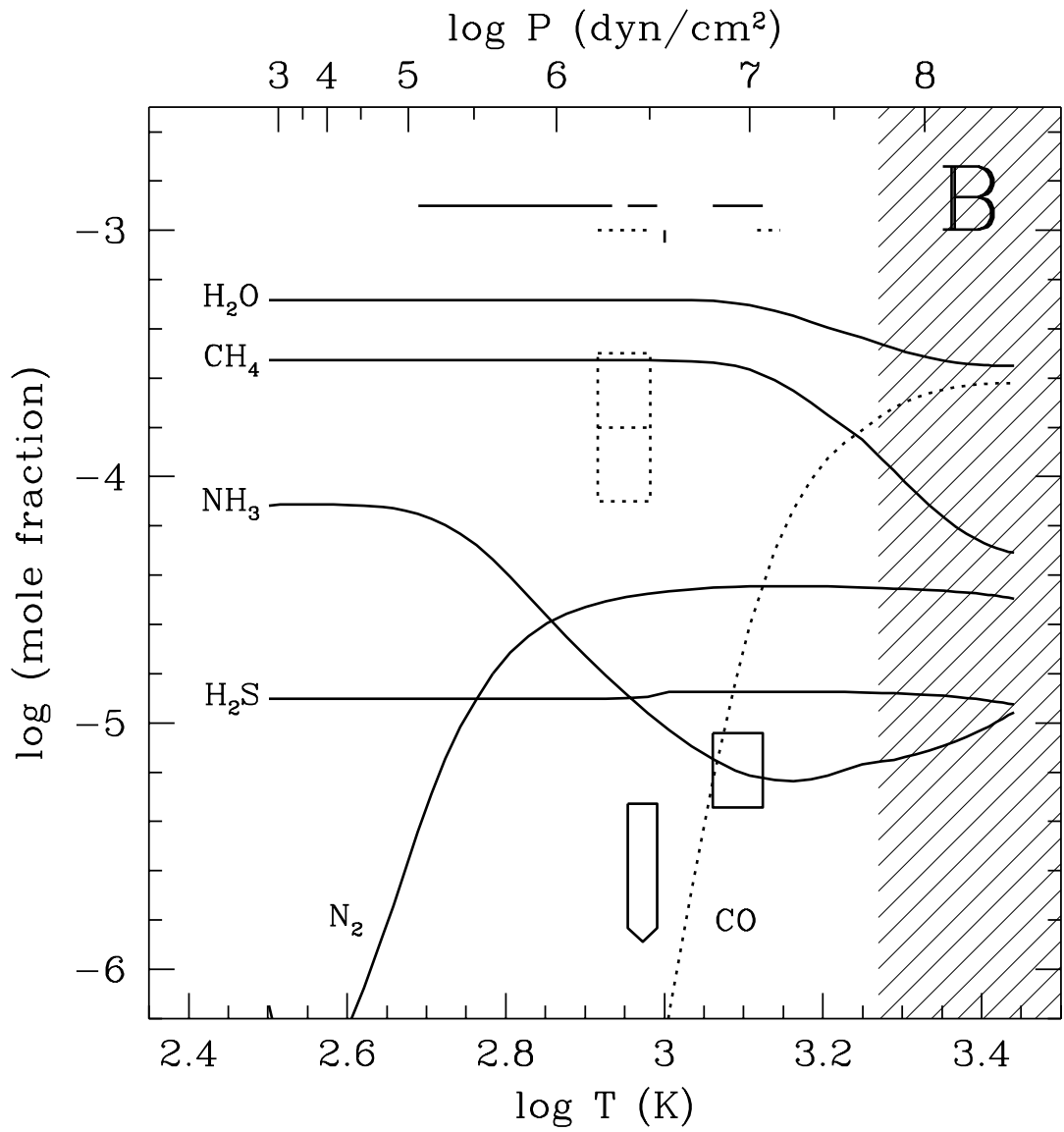
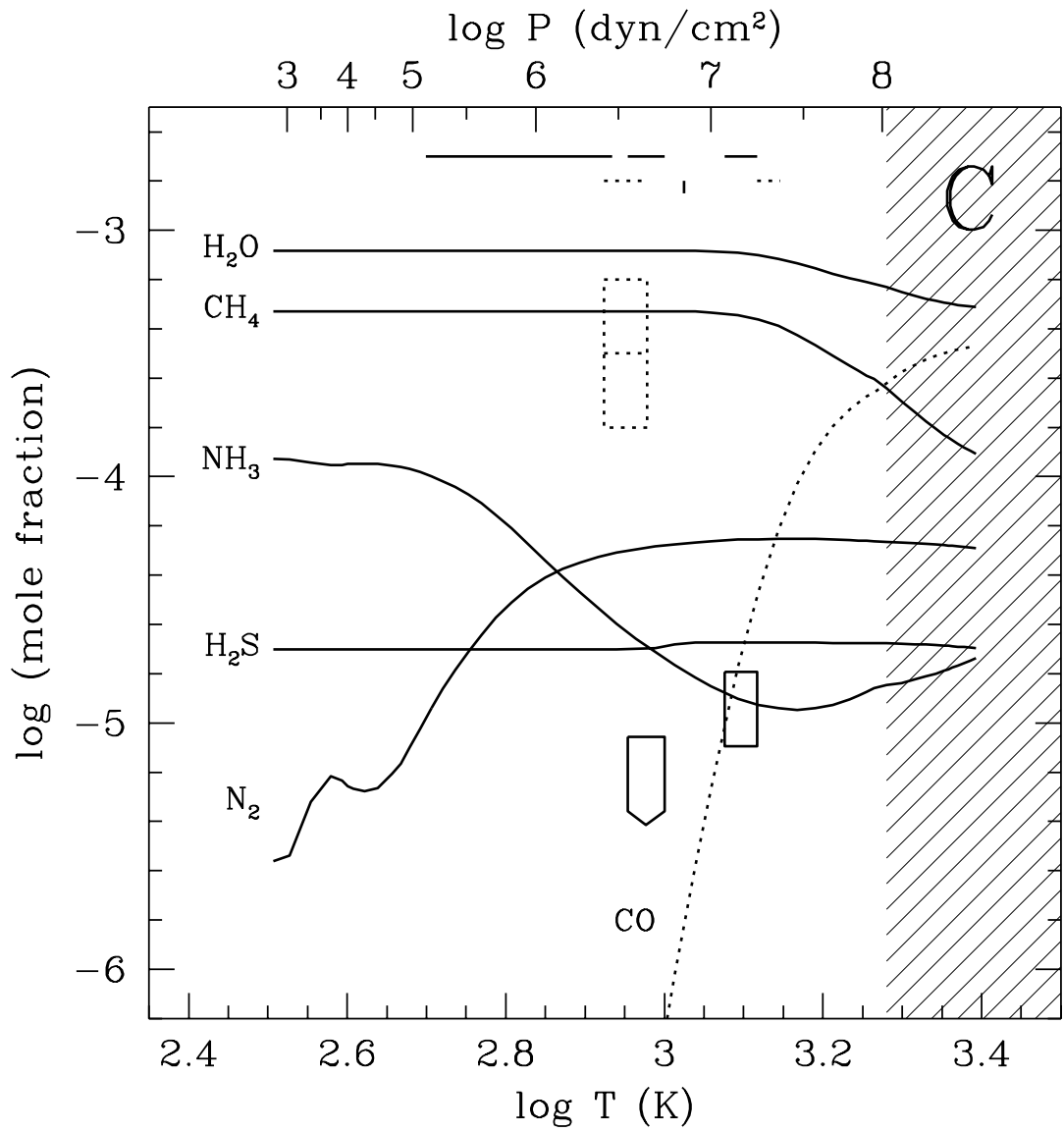
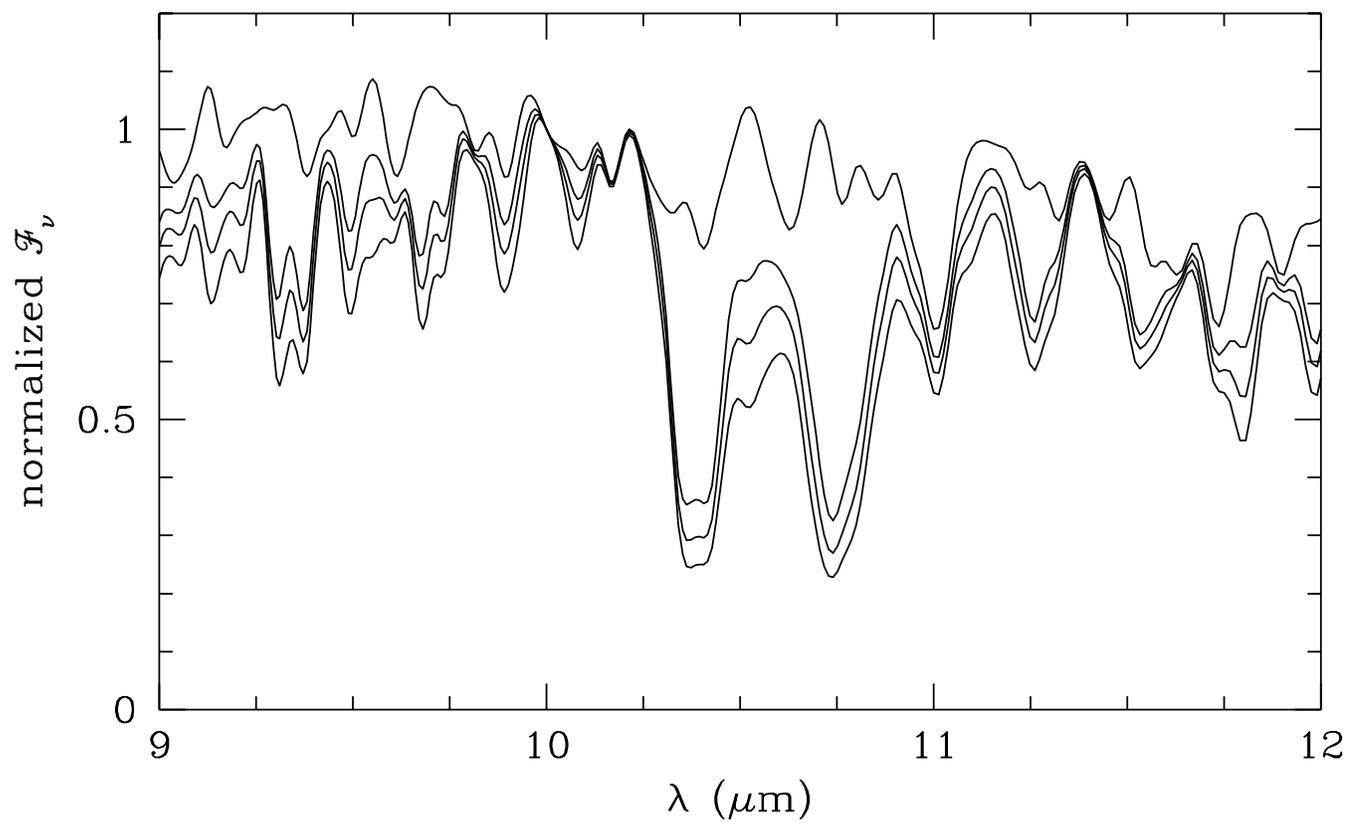
\end{document}